\documentclass[fleqn,usenatbib]{mnras}

\usepackage[T1]{fontenc}
\usepackage{ae,aecompl}

\usepackage{graphicx}	
\usepackage{amsmath}	
\usepackage{amssymb}	
\usepackage{longtable,booktabs,pifont}
\usepackage{enumitem}
\usepackage{color}
\usepackage{newtxtext,newtxmath}

\newcommand{\cmark}{\ding{51}}%
\newcommand{\xmark}{\ding{55}}%

\title[MKN 110 BLR response]{The long-term broad-line responsivity in MKN 110}

\author[D.Homan et al.]
{D.Homan$^{1,2}$, A.Lawrence$^2$, M.Ward$^3$, A.Bruce$^2$, H.Landt$^3$, C.MacLeod$^4$, 
\newauthor M.Elvis$^5$, B.Wilkes$^{5,6}$, J.P.Huchra$^{5,7}$, and B.M.Peterson$^8$ \\
$^1$Leibniz-Institut f\"{u}r Astrophysik Potsdam, An der Sternwarte 16, 14482 Potsdam, Germany \\
$^2$Institute for Astronomy, SUPA (Scottish Universities Physics Alliance), University of Edinburgh, \\ 
Royal Observatory, Blackford Hill, Edinburgh EH9 3HJ, UK \\
$^3$ Department of Physics, Durham University, South Road, Durham DH1 3LE, UK\\
$^4$ BlackSky, 1505 Westlake Ave N \#600, Seattle, WA 98109, USA\\
$^5$ Center for Astrophysics $|$ Harvard \& Smithsonian, 60 Garden St, Cambridge, MA 02138, USA\\
$^6$ HH Wills Physics Laboratory, University of Bristol, Tyndall Avenue, Bristol BS8 1TL, UK\\
$^7$ Deceased, 2010\\
$^8$ Retired
}

\date{Accepted XXX. Received YYY; in original form ZZZ}

\pubyear{2022}

\begin{document}
\label{firstpage}
\pagerange{\pageref{firstpage}--\pageref{lastpage}}
\maketitle

\begin{abstract}
We examine the long-term history of the optical spectrum of the extremely variable Active Galactic Nucleus (AGN) MKN~110. By combining various archival data with new data, we cover an unprecedented long period of $\sim$30~years (1987 -- 2019). 
We find that the He~II~$\lambda 4686$ emission line
changes by a factor of forty and varies more strongly than the optical continuum. Following Ferland et al. (2020), 
we take He~II~$\lambda 4686$ as a proxy for the FUV continuum and compare the flux of several other line species against it. This comparison reveals a clear pattern, whereby lines respond close to linearly at low FUV fluxes, and saturate at high FUV fluxes. 
The saturation level of the response appears to depend on the excitation energy of the line species.
In addition to this global pattern, we note changes among observational epochs, 
indicating a structural evolution in the broad line region (BLR). The line profiles in our spectra show an offset between the narrow and broad components of the He~II~$\lambda 4686$ and H$\beta$ lines. This offset shows a significant negative correlation with the FUV flux and a positive correlation with the line velocity width. Our analysis reveals a complex BLR response to a changing continuum.
The clear presence of a non-responsive component of the broad lines indicates the existence of multiple contributions to the line emission.
We find there are several kinematic models of the BLR and inner regions of the AGN that match our data. 
\end{abstract}

\begin{keywords}
galaxies: Seyfert -- galaxies: nuclei -- quasars: emission lines
\end{keywords}




\section{Introduction}
\label{sec:intro}
Two striking features of the optical/UV spectra of Active Galactic Nuclei (AGN) are the blue continuum and the broad emission lines (BELs). These features are closely connected: they are always seen together, and it is assumed that an extrapolation of the blue continuum into the far-UV provides the photons that ionise the gas producing the emission lines \citep[see e.g.,][and extensive references therein]{Osterbrock-Ferland-book-2006, Peterson_BLR-review-2006, Netzer-book-2013}. Strong evidence for this connection is found in the fact that variations in the strengths of the BELs track the variations in the blue continuum, but with a delay (lag) of the order days-months, depending on the luminosity of the object \citep{Gaskell-Sparke-1986, Clavel1991, Peterson1998, Netzer2020}. The lags, interpreted as light-travel time delays, tell us the size of the Broad Line Region (BLR), using a technique referred to as Reverberation Mapping \citep[RM;][]{Blandford1982, Peterson1993}.

Simple single-cloud photo-ionisation models typically require an ionisation parameter of $U$$\sim$$0.1$~\footnote{$U$ is defined as the ratio of ionising photons to the amount of material that can be ionised: $U = N_{ion}/(4\pi R^2 c N_{atom})$}, along with a typical BLR particle number density of $n$$\sim$$10^{10}\text{ cm}^{-3}$, and a typical cloud thickness $N_H$$\sim$$10^{23}$ $\text{cm}^{-2}$ \citep{Peterson_BLR-review-2006, Netzer-book-2013}. The size of the lag observed in reverberation experiments correlates well with the luminosity of the AGN \citep{Kaspi2000, Bentz2013, Pancoast-reverb-P2-2014, Lira2018, DallaBonta_2020,Netzer2022}, which suggests that the value of $U$$\sim$$0.1$ is quasi-universal. How the universal $U$-value comes about is a long-standing puzzle \citep{Rees-small-1989, Baskin-RPC-P2-PII2013, Baskin2018}, and indeed, although the basic parameters of the BLR - size, density, covering factor - are reasonably well known, its detailed structure, its velocity field, as well as its physical origin, remain controversial \citep{Netzer2020, Netzer2022}.  

A potentially important clue to the structure of the BLR is the \textit{responsivity} of lines to changes in the continuum \citep{Goad1993}. It was already clear in the pioneering study of NGC 5548 \citep{Clavel1991, Krolik1991a} that some lines are more responsive than others. While the UV continuum 1350\AA\ in this object at underwent peak-to-trough changes of a factor of $\sim$$2.5$, line responses ranged from Ly$\alpha$, changing with a factor of $\sim$$1.8$, to Mg~II~$\lambda$2798 with a factor of $\sim$$1.2$. The difference in responsivity may have a natural explanation in terms of the `Local Optimally emitting Cloud (LOC)' picture \citep{Baldwin1995,Korista2000, Korista2004, Goad2015}, as the changing ionising continuum picks out clouds from a broad distribution of density and distance.

However, the response behaviour of lines is complicated. For example, Mg~II is generally assumed to be weakly responsive, and sometimes seems to show no response at all \citep{Cackett2015a,Sun2015}, but extreme variability over several years seen in `Changing Look Quasars' can show large changes in Mg~II~$\lambda$2798 \citep{MacLeod2016,Guo2019,Homan2020}. Meanwhile, the He~II~$\lambda 4686$ line can show extremely strong changes, possibly even larger than the optical continuum \citep{Peterson1986, Peterson1988, Grier2012, Peterson2014, Barth2015, Ferland2020}.

In order to study the BLR response more fully and to clarify these confusing patterns, what is needed are sets of observations that are both high-cadence and long-term, and that involve a large amplitude of variability. Recently, \citet{Ferland2020} have discussed the suitability of using He~II as a proxy for the unobservable FUV continuum since this emission line is effectively a photon counter. Here we explore further the use of He~II as an indicator of the FUV continuum and apply it to the long-term evolution of the broad emission lines in   
the local Seyfert galaxy MKN~110, which has been observed over many years, and has undergone large flux changes.

The paper is organised as follows. In Section \ref{sec:mkn110_history}, we summarise previous work on MKN~110. In Section \ref{sec:data}, we present new spectroscopic observations, and tabulate key values along with data from the literature. We then present our methods for flux calculation and line profile measurements in Section~\ref{sec:fitting}. Section~\ref{sec:history} provides an overview of the long-term evolution of MKN~110 and in Section~\ref{sec:response}, we present an analysis of how various lines respond to the continuum, using He~II~$\lambda 4686$ as a proxy. In Section~\ref{sec:offsets}, we examine how velocity offsets, velocity width, and luminosity state correlate over time. We summarise our results in Section~\ref{sec:keyresult}. In Sections~\ref{sec:discussion-general}--\ref{sec:discussion_kinematic_response} we discuss our new results in the context of rival interpretations of BLR structure and origin, and point towards future work.

\section{MKN~110: key results from previous work}
\label{sec:mkn110_history}

MKN~110 has a long history of observations, focusing on optical, UV, and X-rays \citep[e.g.,][]{Bischoff1999,Ferrarese2001,Boller2007,Porquet2021,Vincentelli2021}. The type E galaxy hosts a type 1 Seyfert nucleus at $z=0.0353$, with an average bolometric nuclear luminosity of $\sim$$2\times 10^{44} \text{erg s}^{-1}$ \citep[][]{VeronCetty2007}. Optical imaging of the galaxy shows a clear tidal tail suggesting a past merger or tidal interaction, but there is only one obvious nucleus. 
The nucleus is clearly detected in X-rays by \textit{ROSAT}, \textit{XMM-Newton}, and \textit{Swift}-BAT with a spectral slope typical of type 1 AGN: for a single power-law fit $\Gamma$$\sim$2.4
\citep{Lawrence1997,Grupe2001,Boller2007,Winter-swift-BAT-2012,Porquet2021}. The nucleus is radio-quiet, observed in NVSS at 10.1 mJy at 1.4GHz \citep{CON92} and observed with the VLBA to be between 0.7 and 1.6 mJy at 5 GHz  \citep[][]{Panessa2022}.

Of particular relevance to this work are several programmes of optical spectroscopic observations \citep[][]{Bischoff1999,Kollatschny2001, Kollatschny2002, Kollatschny2003c,Kollatschny2003b}, that cover the evolution of the continuum and the BLR. \citet[][BK99]{Bischoff1999} collected 24 spectra over eight years (1987 to 1995), measured lags for several lines, and showed that the lags correlate with line width. This is one of the clearest known examples of BLR stratification. \citet[][K01]{Kollatschny2001} undertook a more intense short-term reverberation campaign, with 26 epochs spread over a few months. \citet[][KB02]{Kollatschny2002} and \citet[][K03a]{Kollatschny2003c} used the same data-set, but split several lines into 200 km s$^{-1}$ segments, to produce velocity-resolved delay maps. These maps showed that the line wings have a shorter delay than the cores. The authors argued that the results were consistent with the kinematics of a Keplerian disc. There is marginal evidence that the blue wings lag the red wings by $\sim$2 days, matching a disc-wind model, with a combination of rotation and outflow (K03a). 

\citet[][K03b]{Kollatschny2003b}, again using the K01 data-set, noted that several broad emission lines show a velocity offset towards the red compared to the narrow lines, and that the offset correlates with the line lag, and therefore also with line width -- the offset is especially striking in He~II~$\lambda 4686$. K03b interprets the offset as \textit{gravitational redshift}.
On this basis they derive a central mass of  
log$[M_{BH}/M_\odot] = 8.15^{+0.03}_{-0.10}$. Comparing this with the much smaller K01 reverberation mass, 
log$[M_{BH}/M_\odot] = 7.26^{+0.09}_{-0.11}$, K03b derives an inclination for the BLR disc structure of $i=21\pm 5^\circ$, i.e. close to pole on.

There is a large discrepancy between the K01 and K03b mass estimates, which has a significant impact on the derived physical parameters of the BLR and the classification of MKN~110. Discrepant results for different methods of mass estimation are found in other works as well. As MKN~110 is a well-studied object, there is a large number of independent mass estimates. Based on X-ray data, log$[M_{BH}/M_\odot] = 7.37$ was reported using the excess variance in the flux \citep[][]{Ponti2012}, and log$[M_{BH}/M_\odot]$ in the range 7.4--8.3 using a method based on a Comptonisation model to scale from stellar-mass BHs to SMBHs \citep[][]{Gliozzi2011,Williams2018}. Reverberation mapping studies, including BK99 and K01, reported log$[M_{BH}/M_\odot]$ in the range 6.89--7.78 \citep[BK99,K01][]{Kaspi2000,Peterson2004}. The broad range in the RM values is in large part due to the uncertainty in the virial factor \textit{f}~\footnote{virial factor $f\equiv 1$/(4 $\mathrm{sin}^2 i)$, following \citet{Mclure2004}, who assume a flattened BLR geometry. However, other studies have assumed different BLR geometries resulting in different interpretations of $f$.}, which accounts for the unknown orientation and geometry of the BLR. The assumed values of \textit{f} vary from $\sqrt3/2$ to 5.5 (covering spherical and flattened BLR geometries), among the different studies. Based on the stellar velocity dispersion in the bulge, measured by \citet[][]{Ferrarese2001}, a log$[M_{BH}/M_\odot]$ of $\sim$6.5 can be deduced \citep[][]{VeronCetty2007,Ferrarese2000,Greene2006}. Finally, using spectropolarimetry of the H$\alpha$ line, a method that should be independent from the BLR's inclination, \citet[][]{Afanasiev2019} reported a (log) mass of 8.32$\pm$0.21 M$_\odot$. 

The BELs of MKN~110 are relatively narrow, which means the nucleus could tentatively be identified as a narrow-line Sy1 \citep[e.g.][BK99]{Grupe2004}. However, the X-ray power-law index and low X-ray absorbing column density \citep[][]{Lawrence1997,Porquet2021}, as well as the strong [O~III] lines and weak Fe~II emission \citep[][]{VeronCetty2007} indicate that this classification is likely incorrect. NLSy1s typically have low black-hole masses and accretion rates near or over the Eddington limit \citep[][]{Nicastro2000}, whereas the range of black-hole mass estimates for MKN~110 better matches BLSy1s. \citet[][]{Meyer2011} find an average Eddington ratio for MKN~110 of $\sim$40\% \citep[using the mass from][]{Peterson2004}. The most likely explanation appears to be that MKN~110 is a BLSy1 viewed close to pole-on. 


MKN~110 has long been known as a highly variable source. Optical photometric variability on timescales smaller than days was noted by \citet[][]{Webb2000} and MKN~110 was recently the topic of an extensive continuum-RM campaign \citep[][]{Vincentelli2021,Vincentelli2022}. The continuum RM studies not only showed clear correlations between X-ray, UV, and optical photometric bands on a timescale of days--weeks, but interestingly found that X-ray to UV lags and UVW2 to U-band lags evolve with the luminosity state of the nucleus. In the case of the U-band lag, the authors suggest that the correlation with the luminosity state could be explained by a continuum-dependent contribution by the diffuse BLR gas to the continuum \citep[][]{Vincentelli2022,Goad2019}. In X-rays, MKN~110 shows variability similar to other BLSy1, based on a comparison of their normalised excess variance \citep[][]{Ponti2012,Porquet2021}. Recent VLBA observations also indicate the presence of variable 5 GHz radio emission, on timescales of days--weeks, with the fastest varying emission originating within $\sim$360 gravitational radii of the nucleus \citep[][using a mass of log($M_{BH}/M_\odot$) = $7.68^{+0.15}_{-0.23}$]{Panessa2022}.

The series of spectroscopic RM papers (BK99, K01, K03b) produces an interesting and self-consistent story. The pole-on viewing angle would provide an explanation for the strong discrepancy between the mass estimates based on RM and those independent of the orientation of the BLR \citep[K03b,][]{Liu2017}, as most of the BLR motion would be perpendicular to our line of sight. However, a pole-on view cannot account for the lower mass estimate based on the stellar velocity dispersion \citep[][]{Ferrarese2001}, which matches the RM results. A different potential counter-indication is found in NGC 4051, in which a large He~II velocity offset has been noted \citep{Peterson2000}. In this case, the offset is \textit{to the blue}, which rules out gravitational redshift. Peterson et al. interpret the offset as the sign of an outflowing wind. We argue that although it is not required, it would be preferable if both blue and red offsets could be incorporated into a unified picture. A large part of the analysis presented in this paper will focus on the relation between the BELs and the ionising continuum, however we will return to the discussion of black-hole mass and the nature of the offset of the broad and narrow emission lines in Sections~\ref{sec:offsets} and~\ref{sec:discussion_kinematic_response}. 

\section{data-sets used in this paper}
\label{sec:data}
Our data-set consists of flux measurements from literature and of optical spectra drawn from a range of observing campaigns, including new observations. For those observations for which we have the spectrum available we use spectral fitting to calculate the line fluxes, as well as to extract information on the line profiles. An overview of our data and their provenance is given in Table~\ref{table:datasets_overview}. This table also includes information on the average observational parameters and conditions, where available. In total the sample consists of 207 spectral epochs spread over 28 years (1987 -- 2019). For 62 of the epochs the spectrum itself was used in our analysis. For the other 145 epochs, referred to as the `literature data' below, we only have measured fluxes available.
\begin{table*}
\caption{Overview of our available data on MKN~110. 
}
\label{table:datasets_overview}
\centering
\begin{tabular}{l|c|c|c|c|c|c||l}
\toprule
Data~$^a$ & Epoch & N$_{spec}$ & Spectra~$^b$ & Fluxes~$^c$ & Resolution & Slit/Fibre & Observing conditions \\
\hline\hline
WHP98 & 1992--1996 & 95 & \xmark & \cmark & 1600 (at 4800\AA)& - & - \\
BK99 & 1987--1995 & 24 & \xmark & \cmark & 1300 (at 4000\AA) & 2-2.5" & non-photometric \\
K01 & 1999--2000 & 26 & \xmark & \cmark & 650 (at 4000\AA) & 2" & - \\
SDSS & 2001 & 1 & \cmark & \cmark & 1500 (at 3800\AA) & 2" & photometric \\
FAST & 2002--2019 & 59 & \cmark & \cmark & 1600 (at 4800\AA) & 3" & mixed \\
WHT & 2016--2017 & 2 & \cmark & \cmark & 1500 (at 5200\AA) & 1-1.5" & photometric (2016)/cloudy (2017)\\
\bottomrule
\end{tabular}
\flushleft{\scriptsize{
a) Name of data-set. WHP98 is a previously unpublished data-set of RM observations, BK99 and K01 refer to \citet{Bischoff1999} and \citet[][]{Kollatschny2001}, respectively. FAST and WHT refer to instruments. For the FAST we make us of both archival and new spectra. The WHT data-set contains only new spectra.\\
b) Check mark if the spectra themselves are available.\\
c) Check mark if the line and continuum fluxes are available.\\
}}
\end{table*}

\subsection{Literature data}
The key studies of MKN 110 by BK99 and K01 provide line and continuum fluxes at a total of 49 epochs spread over 1987--2000. In this paper we make use of the BK99 tabulated values for H$\alpha$, H$\beta$, He~I~$\lambda$5876, He~II~$\lambda$4686, and the continuum flux at 5100\AA. The data-set from the K01 paper is limited to one observing season (MJD 51480--51665). For these observations K01 provide continuum fluxes at 5100\AA, as well as at 4265\AA\ and 3750\AA\ for a subset of the epochs, and line fluxes for H$\alpha$, H$\beta$, He~II~$\lambda 4686$, He~I~$\lambda 4471$, and He~I~$\lambda$5876.\\
\indent In addition to these published values, we make use of previously unpublished observations by B. Wilkes, J.P. Huchra, and B. Peterson. These observations were part of a RM campaign made using the FAST spectrograph (see below) in the period 1992--1996 (MJD 48651--50389). This campaign included MKN~110. The fluxes have not been published before now. From this archival FAST data-set we have continuum fluxes at 5100\AA\ available, as well as line fluxes for H$\beta$ and He II $\lambda 4686$. We will refer to this data-set as WHP98.

\subsection{Tillinghast (FAST) spectra}
\label{sec:data_till_spec}
In the period 2002-2019 MKN 110 was observed 59 times by the FAST spectrograph on the Tillinghast 60 inch telescope at Mount Hopkins in Arizona \citep{FAST}. The first of these observations were part of a campaign undertaken by B. Wilkes, J.P. Huchra, and B. Peterson, concentrated into two observing seasons: November 2002--May 2003 and November 2003-May 2004 (MJD 52580--52760 and 52944--53126). These were high-cadence campaigns for RM (Note: this a separate campaign from the WHP98 data). Two further FAST spectra were part of studies reported by \citet{Landt2008,Landt2011}. The FAST spectra taken in 2018 and 2019 were part of our monitoring programme.

\indent The spectra were downloaded from the FAST public archive. The public FAST spectra have been reduced, including wavelength calibration, with the FAST reduction pipeline \citep{FASTpipeline}. For inclusion in this study we flux calibrated the spectra. This calibration was performed using available FAST observations of standard stars. These observations were selected to match the target observation in time as closely as possible. The match in observing date lay within a week to a few weeks. We further applied a grey-shift to the FAST spectra, to correct for a possible offset in the flux calibrations between the FAST spectra and the other spectra in our sample. The correction is based on the assumed constancy of the narrow [O~III]$\lambda$4959 and $\lambda$5007 lines. The FAST spectra were scaled to match the [O~III] fluxes of the SDSS spectrum.

\subsection{SDSS spectrum}
\label{data_sdss_spec}
We use the Sloan Digital Sky Survey (SDSS) spectrum of MKN~110 taken in 2001. This is the only SDSS spectrum publicly available for MKN~110 (SDSS DR16). The SDSS observations and spectral reduction pipeline are discussed in \citet{Gunn2006} and \citet{SDSSI}. No adjustments were made to the calibration of this spectrum. This spectrum is of particular importance in our data-set, as the observation caught MKN~110 in an extremely low state, as we discuss in the following sections.

\subsection{William Herschel Telescope spectra}
We observed MKN~110 on two occasions with the ISIS long-slit spectrograph on the 4.2m William Herschel Telescope (WHT) on La Palma. The 5300 dichroic was used along with the R158B/R300B grating in the red/blue arms respectively, along with the GG495 order sorting filter in the red arm. This setup results in a spectral resolution of \textit{R}$\sim$1500 at 5200\AA\ in the blue and \textit{R}$\sim$1000 at 7200\AA\ in the red for a slit width of 1 arcsec.

\indent The observations were carried out at parallactic angle. Cosmic rays were removed from the 2D spectra using the L.A.Cosmic method \citep{vanDokkum1995}, after which we performed standard wavelength and flux calibrations. The flux calibration was performed with standard stars observed within two hours of the observations and at similar airmass.

\section{Measurements of Fluxes and Line Profiles}
\label{sec:fitting}
\indent For the FAST and WHT spectra we have made new measurements of both fluxes and line profiles. The fluxes were then combined with the fluxes from the BK99, K01, and P98 data-sets, to make one combined flux data-set covering the whole observational history of MKN~110. This data-set is used for the analyses presented in Sections \ref{sec:history} and \ref{sec:response}. The line profiles from the FAST and WHT spectra are used in section~\ref{sec:offsets} to study the BLR's kinematics. 

\subsection{Flux measurements}
\label{sec:fitting_fluxes}
We measure the line fluxes for H$\alpha$, H$\beta$, He~I~$\lambda$5876, He~II~$\lambda$4686, He~I~$\lambda$4471, and the continuum fluxes at 5100\AA, 4652\AA, and 3750\AA. To match the BK99 and K01 data, we use the same method as described in BK99: the continuum flux is calculated as a local average and the line flux is calculated by defining a `pseudo-continuum' at the position of the line. The pseudo-continuum is defined as the linear interpolation between two flux values, one either side of the line. This continuum level is then subtracted from the spectrum and the residuals are summed over a specified window for each line to estimate the line flux. To avoid being overly sensitive to fluctuations in a single wavelength bin, we define each of the two flux values by taking the median flux value over 20 wavelength bins and assigning this value to the central wavelength of the range of bins. Table~\ref{table:pseudo_flux_calc} lists the spectral ranges over which the pseudo-continuum is calculated for the different lines species, as well as the spectral ranges that are summed over to find the line fluxes.\\
\begin{table}
\centering
\caption{Overview of the wavelengths and spectral ranges used to calculate continuum and line flux values, using the pseudo-continuum method. This table follows Table 2 in K01.}
\label{table:pseudo_flux_calc}
\begin{tabular}{l | c c}
	\toprule
	Region of interest & Line Window & Pseudo-Continuum Range \\
	\hline \hline
	H$\alpha$ & 6400--6700\AA & 6260--6780\AA \\
	He~I~$\lambda$5876 & 5785--6025\AA & 5650--6120\AA \\
	H$\beta$ & 4790--4940\AA & 4600--5130\AA \\
	He~II~$\lambda$4686 & 4600-4790\AA & 4600--5130\AA \\
	5100{\AA} continuum & -- & 5130--5140\AA \\
	\bottomrule
\end{tabular}
\end{table}
\indent Our flux measurements are provided in Table~\ref{table:fluxes}. For completeness, we include all available values from the literature data in this table as well. We note that the pseudo-continuum method used to derive these values produces a total line flux, which does not distinguish between broad and narrow line components. In Sections~\ref{sec:flux_corr} and~\ref{sec:response_nonres}, we discuss our method for making corrections 
to the flux. The fluxes in Table~\ref{table:fluxes} all represent the total line flux.\\
\begin{table*}
\centering
\caption{Overview of the flux data used in this paper, combining the results from multiple data-sets.
The full data-set is available online.}
\begin{tabular}{l|c|c|c|c|c|c|c|c|c|c}
    \toprule
    MJD & $F_{5100}$~$^a$ & $F_{4625}$~$^a$ & $F_{3750}$~$^a$ & H$\alpha$~$^b$ & H$\beta$~$^b$ & He~II~$\lambda$4686~$^b$ & He~I~$\lambda$4471~$^b$ & He~I~$\lambda$5876~$^b$ & Data-set \\
    \hline \hline
    46828.0 & 3.14$\pm$0.09 & 3.97$\pm$0.1 & 9.0$\pm$0.2 & 1656.0$\pm$24.0 & 477.9$\pm$9.8 & 108.3$\pm$5.2 & 30.8$\pm$5.2 & 68.7$\pm$5.4 & BK99 \\ 
    47229.0 & 2.57$\pm$0.08 & 2.83$\pm$0.1 & 5.2$\pm$0.2 & 1499.0$\pm$22.0 & 389.3$\pm$8.9 & 58.1$\pm$4.2 & 20.6$\pm$4.4 & 54.7$\pm$4.7 & BK99 \\ 
    47438.0 & 1.77$\pm$0.07 & 1.96$\pm$0.1 & 2.9$\pm$0.4 & 1112.0$\pm$18.0 & 287.3$\pm$7.9 & 37.2$\pm$3.7 & 17.6$\pm$4.2 & 33.1$\pm$3.7 & BK99 \\ 
    47574.0 & 3.06$\pm$0.09 & 3.53$\pm$0.1 & 6.6$\pm$0.4 & 1439.0$\pm$21.0 & 347.0$\pm$8.5 & 60.5$\pm$4.2 & 26.7$\pm$4.9 & 51.1$\pm$4.6 & BK99 \\ 
    47663.0 & 4.43$\pm$0.12 & 5.27$\pm$0.1 & 12.5$\pm$0.4 & 1744.0$\pm$24.0 & 475.1$\pm$9.8 & 159.2$\pm$6.2 & 25.8$\pm$4.8 & 61.8$\pm$5.1 & BK99 \\ 
    47828.0 & 1.98$\pm$0.07 & 1.99$\pm$0.1 & - & 1520.0$\pm$22.0 & 357.9$\pm$8.6 & 30.6$\pm$3.6 & 22.5$\pm$4.6 & 41.7$\pm$4.1 & BK99 \\ 
    48651.8 & 5.68$\pm$0.11 & - & - & - & 545.1$\pm$10.9 & 153.2$\pm$12.3 & - & - & WHP98 \\ 
    48670.9 & 5.40$\pm$0.25 & - & - & - & 581.2$\pm$26.4 & 182.8$\pm$14.6 & - & - & WHP98 \\ 
    48676.8 & 5.04$\pm$0.10 & - & - & - & 526.5$\pm$10.5 & 129.5$\pm$10.4 & - & - & WHP98 \\ 
    48691.9 & 4.31$\pm$0.09 & - & - & - & 529.4$\pm$10.6 & 86.1$\pm$6.9 & - & - & WHP98 \\ 
    \bottomrule
\end{tabular}
\flushleft{\scriptsize{a) units of $10^{15}$ erg cm$^{-2}$ s$^{-1}$ \AA$^{-1}$.\\b) $10^{15}$ erg cm$^{-2}$ s$^{-1}$.}}
\label{table:fluxes}
\end{table*}

\subsection{Line Profiles}
\label{sec:fitting_profiles}
For the 62 epochs with available spectra (Table~\ref{table:datasets_overview}) we perform spectral fitting to extract detailed information about the observed changes in line profiles. We derive line widths and central wavelengths for different line components in He~II~4686 and H$\beta$. The fitting is performed using the \texttt{lmfit} package\footnote{https://lmfit.github.io/lmfit-py/}. For the Tillinghast spectra taken between MJD 52252 and 52798 and between MJD 52940 and 53140 the He~II flux is so low that any information on the line profiles in the individual spectra is lost in the noise. In order to nonetheless extract information from this data-set, we stacked these spectra and fitted the average `low-state' spectrum. Those spectra with sufficient He~II $\lambda$4686 flux for a successful fit were modelled individually.
 
Through experimentation with the fitting window we find the optimal procedure is to fit the He~II and H$\beta$ lines separately, in separate windows. The wavelength ranges for the windows are selected to split the problematic overlap region between He~II and H$\beta$. For He~II, the fitting region is set to 4400--4810\AA\ (QSO restframe), and for H$\beta$ it is set to 4820--5050\AA. Next, we test the effects of including a range of spectral model components: Gaussian or Lorentzian line profiles for He~II, H$\beta$, [O~III] $\lambda$4959, and [O~III] $\lambda$5007, additional narrow line components for He~II and H$\beta$, a power-law continuum of the form $A\nu^{-\alpha}$, and an optical Fe~II emission template \citep{VeronCetty2004}, convolved with a Gaussian kernel of variable width. 

\begin{figure}
\begin{minipage}{.49\textwidth}
\centering
\includegraphics[width=.8\textwidth]{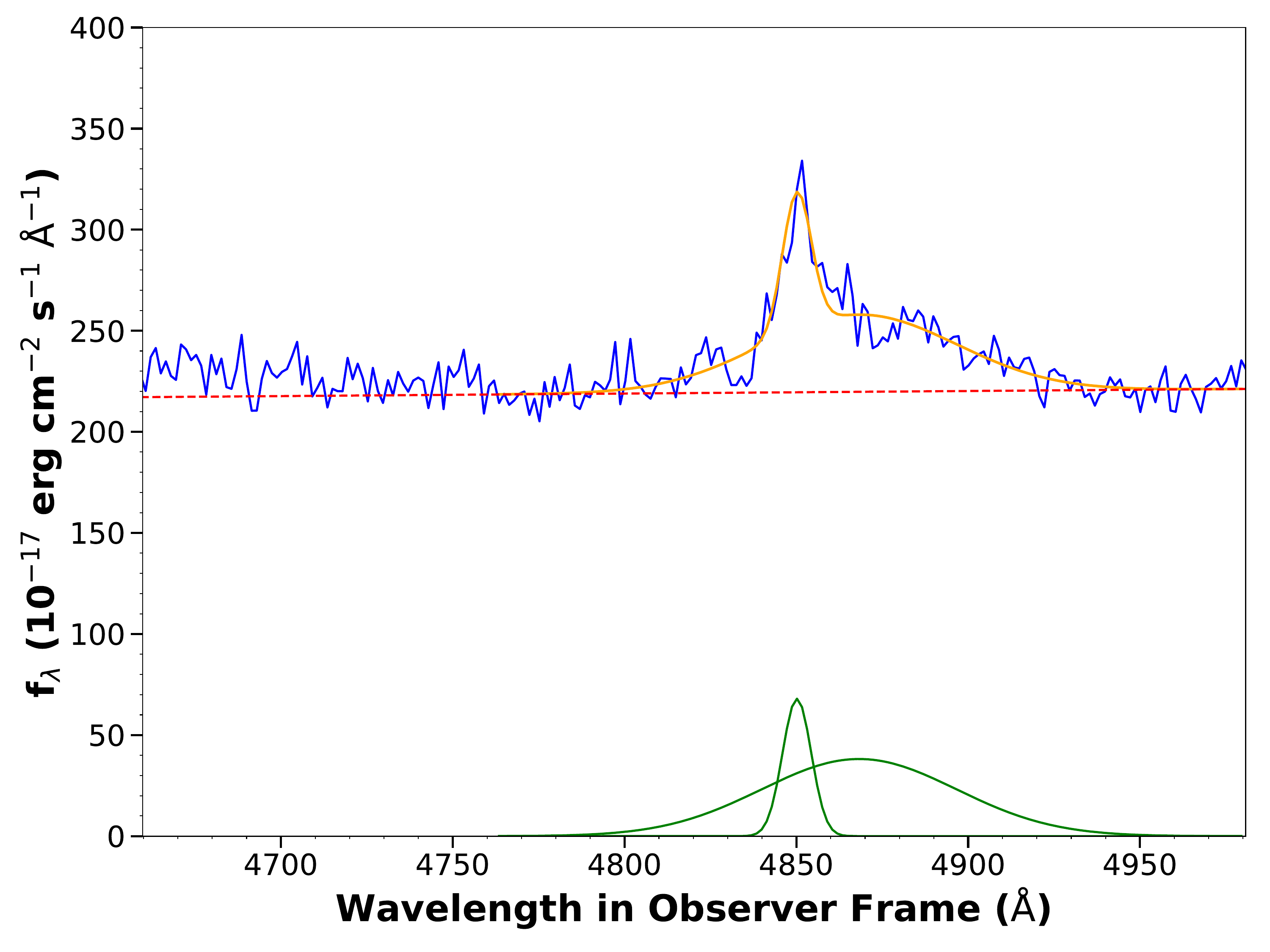}
\end{minipage}
\begin{minipage}{.49\textwidth}
\centering
\includegraphics[width=.8\textwidth]{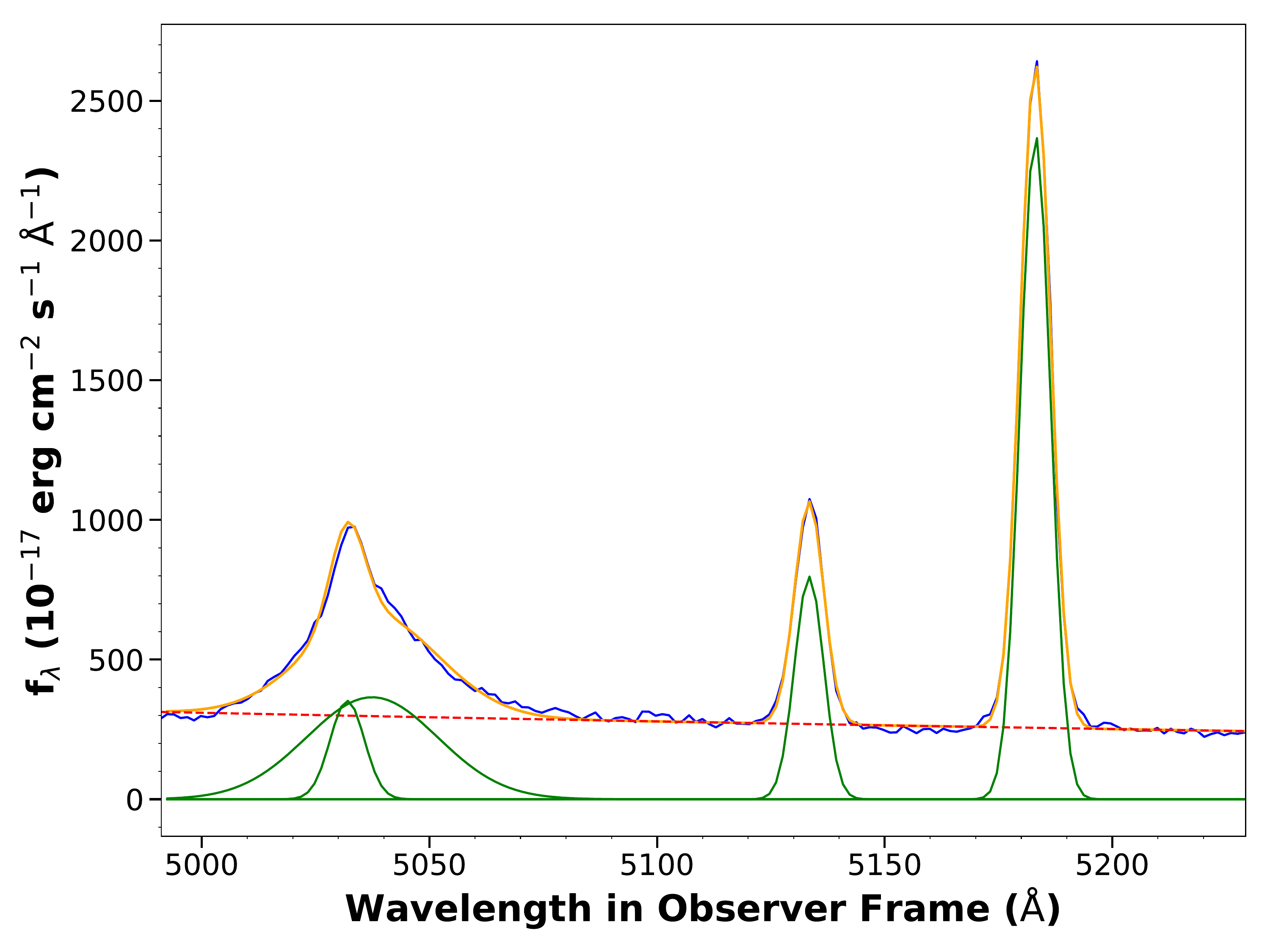}
\end{minipage}
\caption{Example of the results of our fitting method for the two line species, as described in the text. The spectra were both taken with FAST. \emph{Top}: The He~II~$\lambda$4686 line fit (orange) containing the power law continuum (dashed red) and two Gaussian components (green). The spectrum was taken on MJD 52997. \emph{Bottom}: The H$\beta$ line fit (colours as for He~II), with power law continuum and Gaussian line fits to narrow H$\beta$, broad H$\beta$, and the two [O~III] lines. The spectrum was taken on MJD 53021.}
\label{fig:fit_example}
\end{figure}

Our final fitting model is deliberately minimal. All many-component models had a strong tendency towards over-fitting. An example of the results of our final fitting procedure for the two line species is shown in Figure~\ref{fig:fit_example}. The components included in the model for the He~II region are the continuum power law and two Gaussians, for the broad and the narrow component of the line respectively. The amplitude, width, and centre for the narrow line are fixed. The narrow line parameters are set to those derived in a free fit to the spectrum of MJD 53019, a high S/N FAST spectrum. The line flux and central wavelength of the He~II narrow line in MKN~110 are therefore assumed to be constant during the time spanned by these observations, which is approximately fifteen years. For H$\beta$, the model consists of the same power law and four Gaussian components. The Gaussian components are used to model broad and narrow H$\beta$, [O~III]$\lambda 4959$, and [O~III]$\lambda 5007$. We list our fitting results for He~II and H$\beta$ in Table~\ref{table:spec_fit_results}.

\begin{table*}
    \centering
    \caption{The results of our spectral fitting of the He~II~$\lambda$4686 and H$\beta$ lines. The full data-set is available online.
    }
    \begin{tabular}{c|ccc|ccc|c}
    \toprule
    MJD & \multicolumn{3}{c}{He~II~$\lambda$4686$_{broad}$} & \multicolumn{3}{c}{H$\beta_{broad}$} & \multicolumn{1}{c}{H$\beta_{narrow}$}\\
     & $\lambda_c$~$^a$ & flux~$^b$ & $\sigma$~$^a$ & $\lambda_c$~$^a$ & flux~$^b$ & $\sigma$~$^a$ & $\lambda_c$~$^a$ \\
    \hline \hline
    52961 & 4878.21$\pm$2.39 &  19.12$\pm$2.19 &   28.07$\pm$2.80 & 5040.26$\pm$0.51 &  89.35$\pm$4.14 & 16.27$\pm$0.64 & 5032.21$\pm$0.03 \\
    52975 & 4872.73$\pm$1.74 &  25.65$\pm$2.03 &  29.31$\pm$2.03 & 5038.91$\pm$0.39 & 110.57$\pm$3.96 & 14.97$\pm$0.47 & 5031.97$\pm$0.03 \\
    52988 & 4873.81$\pm$9.15 & 31.08$\pm$13.23 & 43.79$\pm$12.75 &  5037.80$\pm$0.49 &  97.56$\pm$4.73 & 13.67$\pm$0.58 & 5031.99$\pm$0.04 \\
    52997 & 4869.03$\pm$1.59 &  24.95$\pm$1.85 &   26.71$\pm$1.80 &  5038.53$\pm$0.30 & 137.97$\pm$3.78 & 14.26$\pm$0.35 & 5032.59$\pm$0.04 \\
    53001 & 4866.63$\pm$2.22 &  24.98$\pm$2.39 &   31.90$\pm$2.63 & 5037.66$\pm$0.26 & 154.69$\pm$3.75 & 14.07$\pm$0.31 & 5032.04$\pm$0.03 \\
    53005 & 4866.26$\pm$2.19 &  18.02$\pm$1.99 &  22.97$\pm$2.41 & 5037.25$\pm$0.26 & 151.16$\pm$3.67 & 14.52$\pm$0.31 & 5031.76$\pm$0.03 \\
    53016 & 4873.31$\pm$4.87 &    7.33$\pm$2.00 &  20.45$\pm$5.38 &  5037.53$\pm$0.30 & 138.93$\pm$3.96 & 14.74$\pm$0.37 &  5031.60$\pm$0.03 \\
    53019 &  4872.98$\pm$2.30 &  14.23$\pm$1.63 &  24.12$\pm$2.56 & 5036.64$\pm$0.25 & 142.73$\pm$3.32 &  14.01$\pm$0.30 & 5031.56$\pm$0.03 \\
    53021 & 4873.52$\pm$2.47 &  24.07$\pm$2.65 &  27.64$\pm$2.92 & 5037.42$\pm$0.29 & 131.99$\pm$3.65 & 14.42$\pm$0.35 & 5031.96$\pm$0.03 \\
    53047 & 4869.94$\pm$2.43 &   21.10$\pm$2.46 &   25.40$\pm$2.73 & 5037.78$\pm$0.27 & 167.45$\pm$4.19 & 14.53$\pm$0.33 & 5032.55$\pm$0.03 \\
    53049 & 4868.07$\pm$2.76 &  27.03$\pm$3.22 &  33.32$\pm$3.34 & 5037.27$\pm$0.27 &  172.79$\pm$4.40 & 14.44$\pm$0.33 & 5031.99$\pm$0.03 \\
        \bottomrule
    \end{tabular}
    \flushleft{\scriptsize{
    a) units of \AA.\\
    b) units of $10^{15}$ erg cm$^{-2}$ s$^{-1}$.
    }}
    \label{table:spec_fit_results}
\end{table*}
\section{History of MKN~110 variability}
\label{sec:history}
\subsection{Evolution Over Three Decades}
\begin{figure*}
\centering
\includegraphics[width=.8\textwidth]{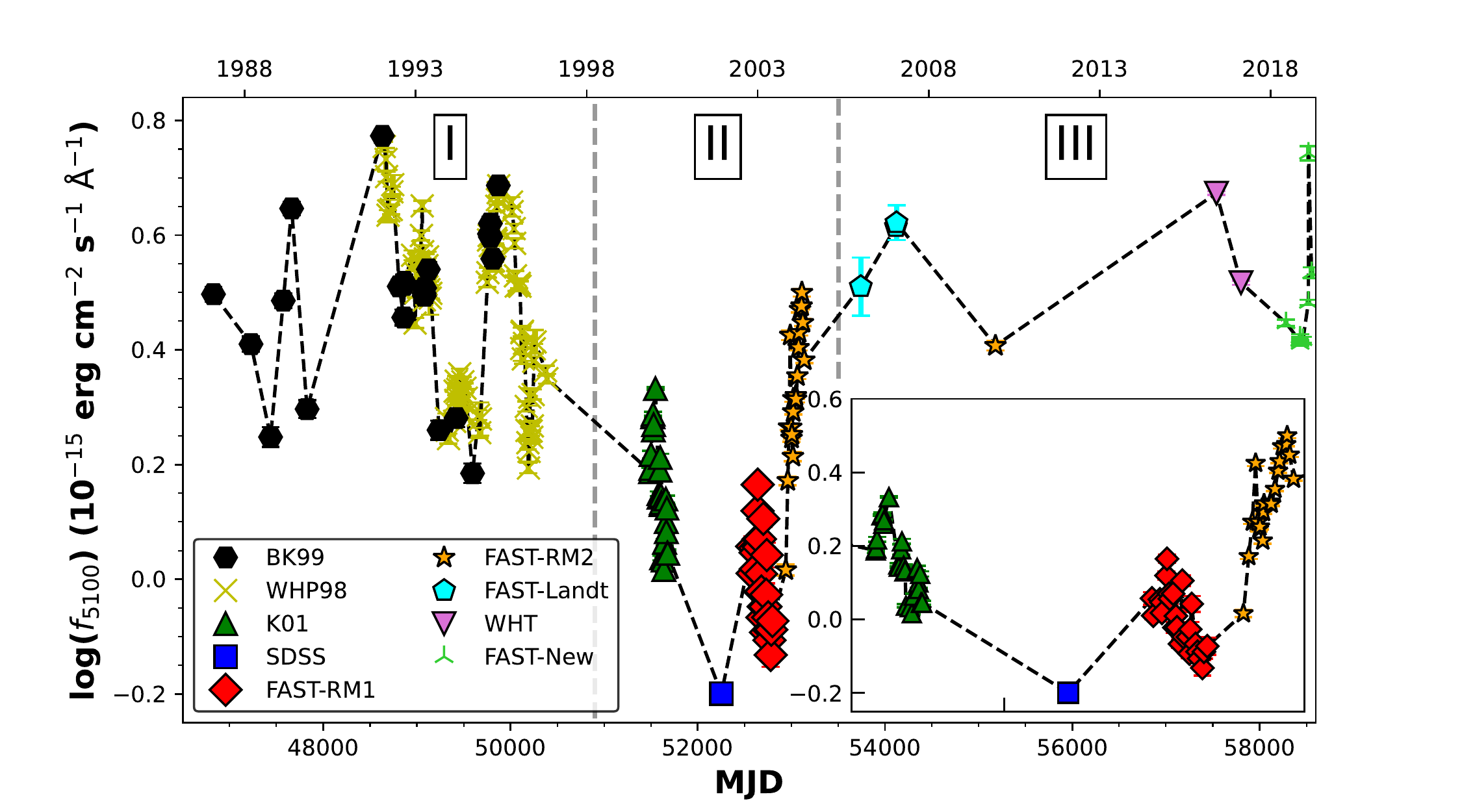}
\caption{The long-term continuum light curve for MKN~110 based on spectrophotometric fluxes at 5100{\AA} ($f_{5100}$). The AGN shows significant variability in its output, including multi-year trends, as well as seemingly more erratic jumps. MKN~110 reached a low state around the time of the SDSS spectral observation. The epochs have been labelled according to the origin of the spectral data, with the abbreviations as listed in Table~\ref{table:datasets_overview}. The archival FAST spectra have been divided into RM1 and RM2, representing two separate Reverberation Mapping campaigns. The Roman numerals and dashed grey lines indicate the three observational epochs we define in our data-set (see Section~\ref{sec:response_hysteresis}). The inset shows a closer view of the observational epochs for the high cadence K01, FAST-RM1 and FAST-RM2 data. These RM data show that MKN~110 continued to display low-level variability in addition to 
the larger structural flux changes.}
\label{fig:lcf5100}
\end{figure*}

Our data-set includes 207 epochs spread over 28 years and contains high-cadence and low-cadence periods of observation. Over this extended period of observation the flux at 5100\AA\ has shown variation by up to a factor of 9.4. We plot the long-term history of the optical continuum in Figure~\ref{fig:lcf5100}. Over large parts of the light curve the variability appears stochastic, as would be expected for AGN \citep[e.g.][]{Kelly2009,Kozlowski2010}. However, our sampling over much of the light curve is too sparse to capture shorter-term changes. The inset in Figure~\ref{fig:lcf5100} shows a detailed look at the three high-cadence campaigns that took place over MJD 51544--53300. We have split the archival FAST spectra into the groupings FAST-RM1 and FAST-RM2, to mark the separate Reverberation Mapping campaigns. From the data in the inset we see that MKN~110 can undergo variations in flux up to a factor two during a single year. Over a single observing season the data show rising and falling trends on a timescale of months.

\begin{figure*}
\centering
\includegraphics[width=0.85\textwidth,angle=0]{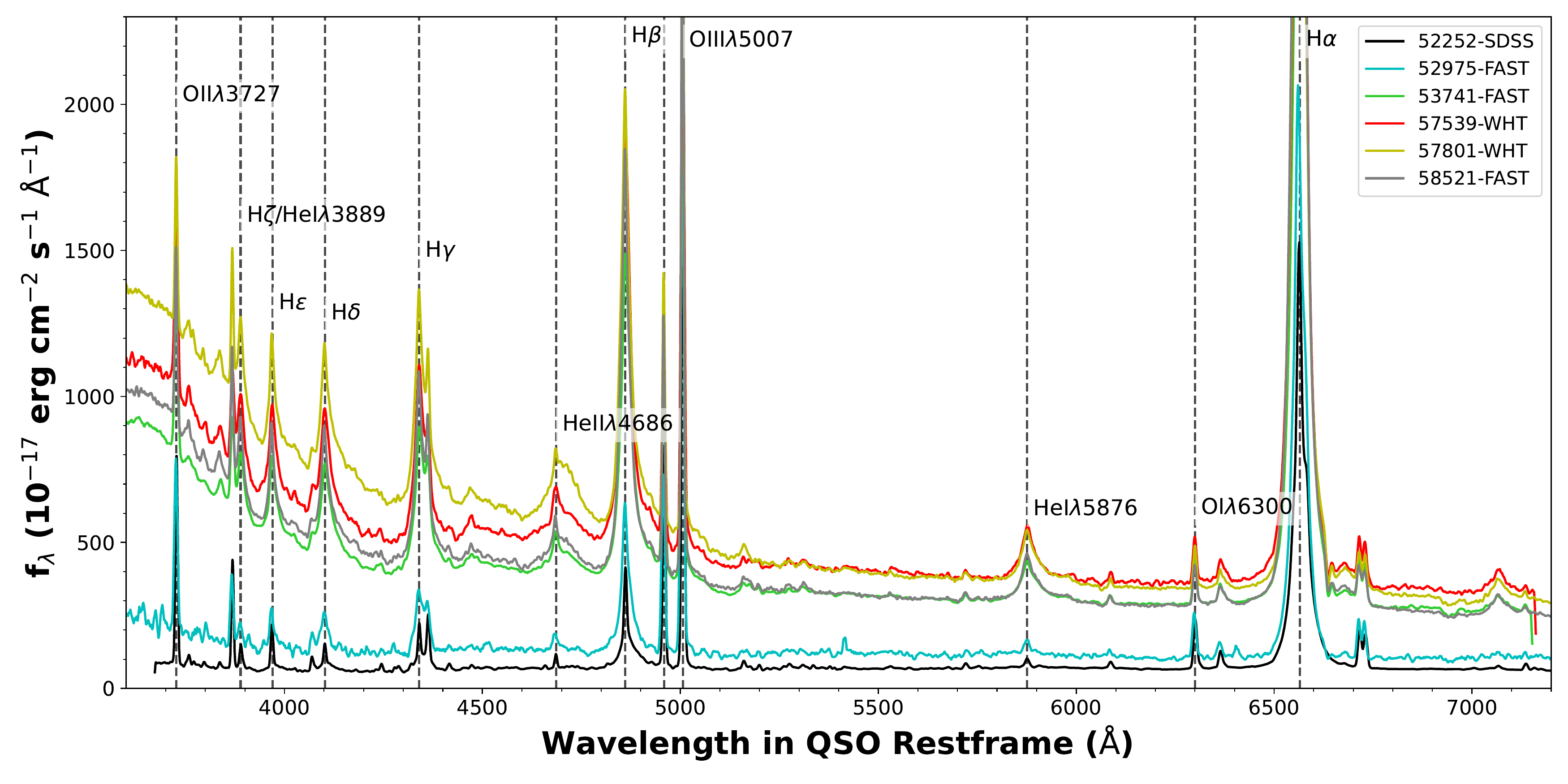}
\caption{A selection of redshift-corrected spectra from high, low, and intermediate luminosity states of MKN~110. The spectra show an evolving blue continuum as well as strong variability in the BELs. We have marked some of the most important emission lines in the plot. Particularly noteworthy is the variability in the broad He~II~$\lambda$4686 line, which is not detected in the SDSS spectrum, is very strong and broad in the WHT spectrum of MJD 57539, and is in intermediate states at other epochs.}
\label{fig:example-spectra}
\end{figure*}

MKN~110's strong photometric variability is accompanied by large changes in its optical spectrum. Figure~\ref{fig:example-spectra} shows a selection of spectra that cover the range of luminosity states. Over the course of the past decades, MKN~110 has shown large changes in its broad emission lines, in particular the broad Helium and Balmer lines. A more distant AGN, observed with lower signal-to-noise, showing the same variations might well have been labelled a Changing Look Quasar \citep[e.g.][]{LaMassa2015}. The spectra also show that MKN~110 gets bluer when brighter, as is typical in AGN \citep[][]{Paltani1994}.
It is clear that while the BELs in MKN~110 change, they do not behave identically. The most striking aspect of the evolution visible in Figure~\ref{fig:example-spectra} is the extremely strong changes in He~II~$\lambda 4686$. In the lowest state (SDSS), broad He~II is undetected. If we compare the next lowest state available to the maximum flux level, He~II changes by a factor of 38. We also calculate the fractional variability \citep[$F_{var}$; cf.][]{Vaughan_2003} over the full time range of our data. We find $F_{var}$ of 0.460$\pm$0.002 for the 5100{\AA} continuum flux, 0.338$\pm$0.001 for H$\alpha$, 0.367$\pm$0.002 for H$\beta$, and 0.694$\pm$0.006 and 0.698$\pm$0.020 for He~II~$\lambda$4686 and He~I~$\lambda$5876, respectively. We will investigate this remarkable differential behaviour of the various line species. In the rest of this section we detail how we prepare our data-set for this analysis.

\subsection{Corrections to the Flux Data}\label{sec:flux_corr}
\noindent To present a uniform data-set, we make a number of corrections to our flux data. The first is to compensate for the narrow-line contribution to the He~II~$\lambda$4686 fluxes. The pseudo-continuum method we have used to measure the line fluxes does not distinguish between the broad and narrow components of a line. It simply returns all the emission above the continuum level in a given interval (Section~\ref{sec:fitting_fluxes}). As we wish to investigate the behaviour of the broad emission lines, we require a correction for the emission from the Narrow Line Region (NLR). Although it is possible to model the broad and narrow components separately through spectral fitting, we do not have the spectra available for the literature data-sets. We therefore require a more general method of correction. Visual inspection of the He~II~$\lambda 4686$ line (Figure~\ref{fig:example-spectra}) shows that the He~II narrow component is quite distinct. The low-state SDSS spectrum, in which the broad component has vanished, contains a clear narrow He~II component. 

We therefore assume the narrow He~II component to be constant over our period of investigation. We note there is evidence for variability of the [OIII] narrow lines in NGC~5548 on a timescale of decades, significantly longer than the broad line response timescale \citep{Peterson2013}. We are not able to exclude such narrow-line variability from our data-set, which may therefore introduce an additional uncertainty in the He~II broad component fluxes. This is taken into account when we estimate the total uncertainty in the line fluxes in Section~\ref{sec:response_quant}. We measure the flux of narrow He~II~$\lambda 4686$ in the SDSS spectrum, fitting a single Gaussian to the line. Our correction to the He~II flux data consists of subtracting the flux of the narrow component in all epochs, giving us an estimate of the broad-line flux. For the other line species (H$\alpha$, H$\beta$, and He~I~$\lambda$5876) there is no epoch in which the broad component is completely gone. We will correct these fluxes using a different method than for He~II. As this other method requires us to first evaluate the response of the various line species to the changing luminosity state, we will cover this step later on, in Section~\ref{sec:response_nonres}.

We next adjust the data to account for the BLR's spatial stratification. RM results show that a fluctuation in the FUV flux originating in the central engine will reach the line-forming regions of different line species at different times. Representing the fluxes from the same observational epoch as synchronous measurements implicitly assumes that the line-forming regions are affected by the same ionising flux. This approach would therefore ignore light travel times between the different regions of the stratified BLR. In the subset of our data-set consisting of fluxes from RM studies (K01, FAST-RM1, and FAST-RM2) we are able to to correct for this effect, as the cadence of observations is sufficiently high.

We make use of the lags reported in K01 (Table~\ref{table:resp_fit_results}). The date of each flux observation is shifted by the appropriate lag, after which the fluxes are matched with the closest, lag-corrected 5100\AA\ continuum fluxes. We performed all analyses presented in this paper with the lag-corrected and the original data-set and find no significant changes in our conclusions. We are therefore confident that the inability to correct for lags among the lower-cadence data-sets has no significant impact on our results. 

In the final step of our corrections, the fluxes are normalised. We normalise to the fluxes from a single epoch. The normalisation allows for the investigation of the relative change in line-flux. The fluxes used for the normalisation are those from the observation on MJD 47574, an epoch from the BK99 data-set~\footnote{The choice of date is arbitrary for our normalisation.}. All fluxes at this epoch therefore have a value of unity in our data-set, while fluxes at other epochs are scaled relative to the flux 
on this date. 

To summarise the various changes we make to the data-set: we have corrected the He~II~$\lambda$4686 fluxes by fitting the flux of the narrow component (assumed to be constant) and subtracting it. We have adjusted (where possible) the combination of line-flux and continuum-flux measurements, to compensate for the light-travel time between different line forming regions in the BLR. And finally, we normalised all flux values to the measurements on one particular date: MJD 47574.

\subsection{The connection between emission lines and continuum flux}
\begin{figure}
\centering
\hspace{-.7cm}
\includegraphics[width=1.05\columnwidth]{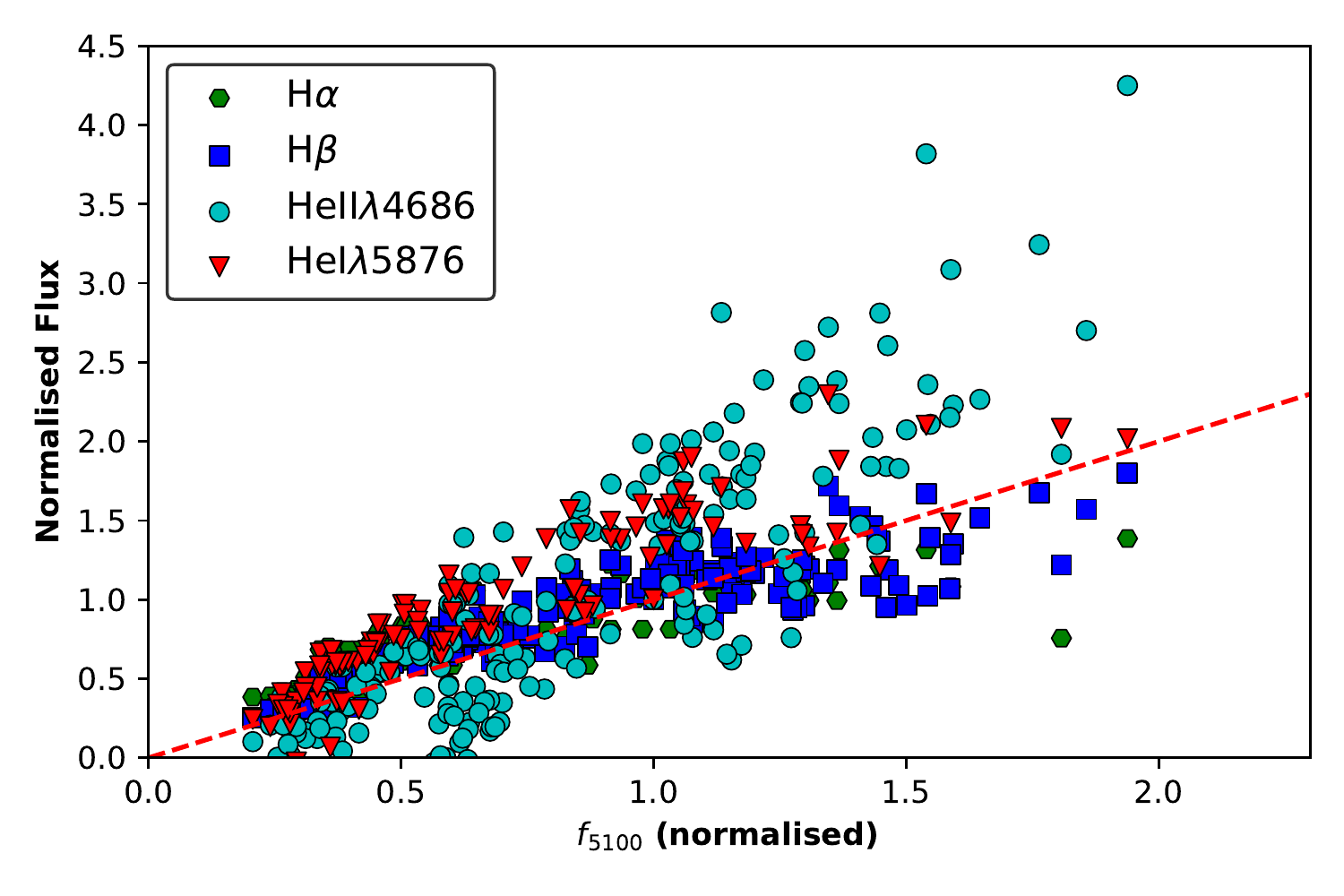}
\caption{The fluxes of four broad-line species in MKN~110. The line fluxes are plotted against the continuum flux at 5100{\AA}. The line-species are H$\alpha$ (green hexagons), H$\beta$ (dark blue squares), He~II $\lambda 4686$ (cyan circles), and He~I $\lambda$5876 (red triangles). All fluxes have been adjusted as specified in Section~\ref{sec:history}. The red dashed line indicates a 1:1 ratio between the line and continuum fluxes. The He~II $\lambda 4686$ flux has a distinct responsiveness compared to the other lines: it increases more rapidly than the optical continuum and continues to grow in flux at higher f$_{5100}$, whereas the response of the other line species appears to flatten.}
\label{fig:mrk110_allfluxes}
\end{figure}

The corrected and normalised fluxes are shown in Figure~\ref{fig:mrk110_allfluxes}, where we plot the line fluxes against the optical continuum at 5100{\AA}. The 5100{\AA} flux is often used in reverberation studies as a proxy for the driving continuum, as this region of the spectrum is relatively free of contamination by blended lines and is more readily accessible to observations than the UV continuum \citep{Peterson_1992}. All line species in Figure~\ref{fig:mrk110_allfluxes} show a clear flux correlation with the continuum, as would be expected. As the optical continuum flux increases, the ionising flux powering the BLR increases as well. We also note a strong differentiation in response among the line species, with the Balmer lines having the flattest response, He~II $\lambda$4686 varying \textit{with a greater amplitude than the optical continuum}, and He~I $\lambda$5876 being intermediate. The slope of the response of the Balmer lines is close to 1:1 with the optical continuum (as indicated by the dashed red line) and shows signs of an even lower response at higher continuum fluxes. In contrast, the He~II line shows no sign of flattening off at high luminosity states.

\subsection{Ruling out reddening}
\begin{figure*}
\hspace*{-1cm}\begin{minipage}{.49\textwidth}
\includegraphics[width=\textwidth]{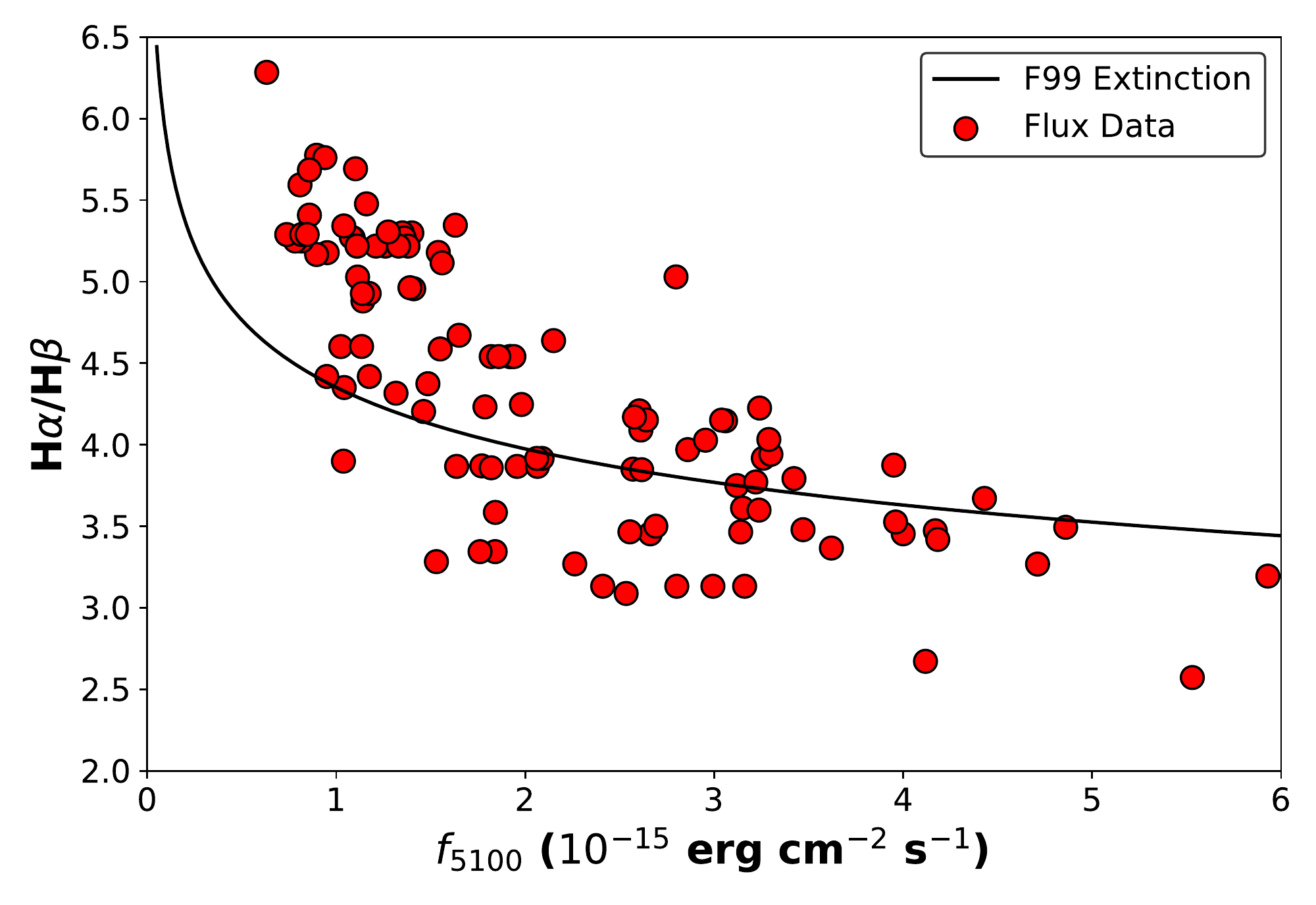}
\end{minipage}
\begin{minipage}{.49\textwidth}
\includegraphics[width=\textwidth]{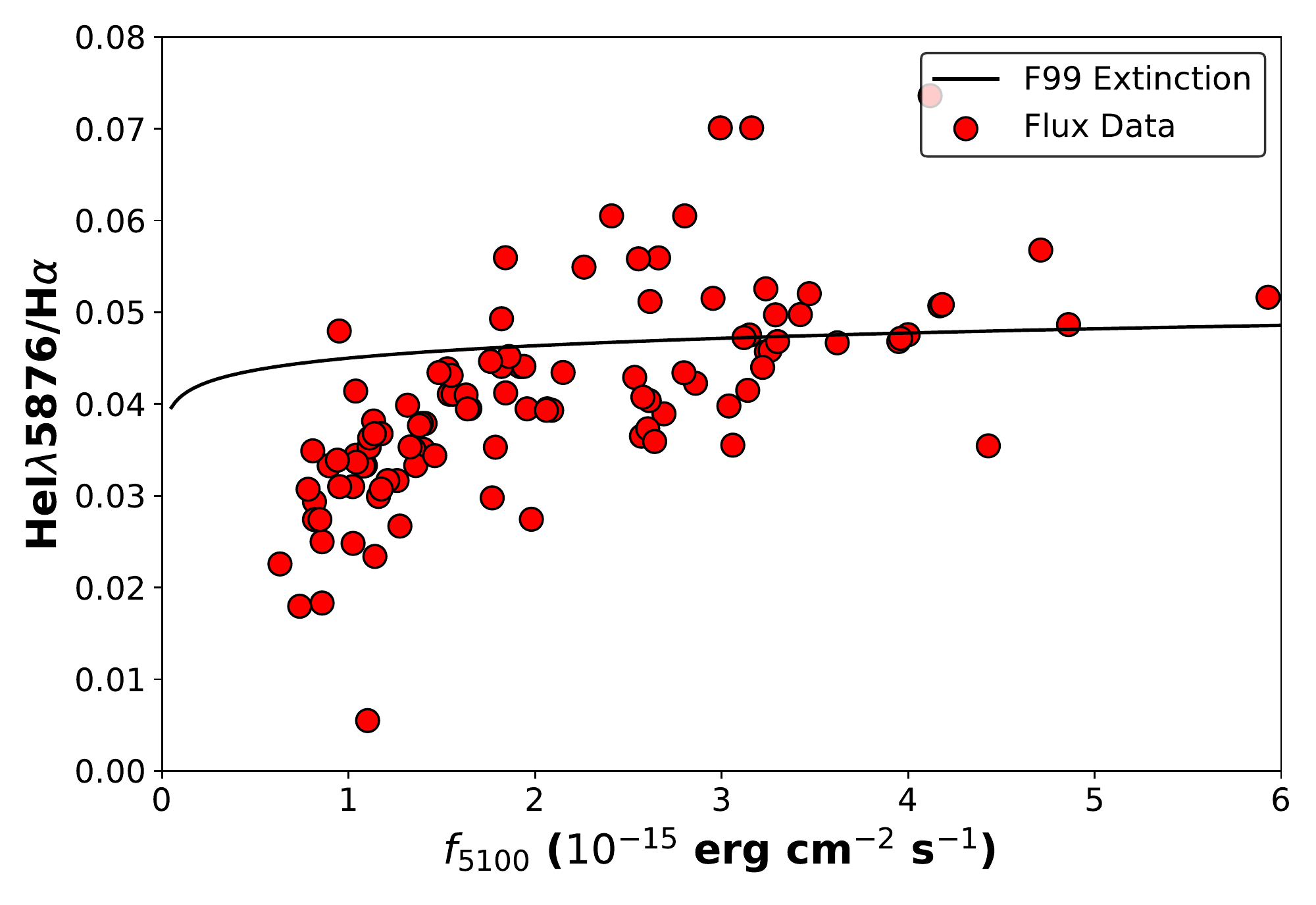}
\end{minipage}
\caption{The line flux ratios H$\alpha$/H$\beta$ ($left$) and He~I~$\lambda$5876/H$\alpha$ ($right$) are plotted against the 5100{\AA} continuum flux. We use these metrics to investigate whether variable obscuration can explain the spectral evolution observed in MKN~110. The theoretical reddening curve is calculated using the extinction curve from \citet[][ F99]{Fitzpatrick1999} under the assumption that all changes in continuum flux are due to dust reddening. The data and theoretical extinction are a poor match. We therefore rule out a changing attenuation by dust as the driving force for the observed variability in the line and continuum flux levels. It appears that these changes are instead driven by changes in the ionising power source.}
\label{fig:mrk110_reddening}
\end{figure*}

We consider whether the detected flux variations can be explained by variable obscuration by intervening dust.
To check for reddening we use the line flux ratios H$\alpha$/H$\beta$ and He~I~$\lambda$5876/H$\alpha$, viewed against the 5100{\AA} continuum flux. 
The dust is modelled as a single sheet and we make use of the extinction curve from \citet[][]{Fitzpatrick1999}. We calculate the level of reddening required to explain the relative change in the 5100\AA\ flux and model the expected value for the given line ratios. The comparison is shown in Figure~\ref{fig:mrk110_reddening}. The reddening modelled for H$\alpha$/H$\beta$ (\emph{left}) shows the same overall behaviour as the data, but still provides a poor fit: it does not match either the relative constancy of the ratio at high luminosities, or the steep rate of change at low luminosities. The ratio of He~I/H$\alpha$ (\emph{right}) shows an even poorer fit. 
Thus, we find that variable, single-screen extinction due to dust does not dominate the variability in line and continuum. It therefore appears likely that the variability in the line fluxes is indeed attributable to changes in the ionising flux, and subsequent reprocessing in the BLR.

\section{The Response of the BLR in MKN~110}
\label{sec:response}
\subsection{Using He~II~$\lambda$4686 as an FUV continuum proxy}
The Balmer and Helium emission lines discussed in this paper are all recombination lines, powered by an ionising (E$>$13.6 eV; FUV) continuum, which we cannot observe directly. Based on the strength of emission lines compared to the observed continuum, it has long been suspected that the continuum powering the broad-line emission, the emission the BLR `sees', is different from the continuum that we can observe directly \citep{Mathews1987,Korista1997}. This has motivated observational studies to approximate the unseen part of the SED, using broad emission lines \citep{Mathews1987,Melendez2011,Panda2019,Ferland2020}. Modelling of AGN SEDs, bound by the constraints provided by observational studies, also shows a limited connection between the optical continuum ($f_{5100}$) and the FUV~\citep{Done2012,Jin2012,Kubota2018}. As the optical region of the SED lies significantly redward of the peak in the spectrum, the FUV flux can take on a wide range of values, with only marginal changes in the optical flux \citep[][]{Kubota2018,Ferland2020}. Reverberation studies have also observed a negative correlation between wavelength and variability amplitude \citep[e.g.,][]{Fausnaugh2016}, which is evident in MKN~110 \citep{Vincentelli2021}. The $f_{5100}$ flux is therefore a useful measure to detect \textit{when} the FUV continuum changes, however it is less suitable to measure by \textit{how much}. For the latter, the broad emission lines provide a better approach.

The He~III recombination lines in AGN, such as He~II~$\lambda$4686, are powered by photons with an energy greater than 54.4~eV. This means He~II lines provide a measure of the continuum flux in the FUV part of the SED, even beyond the H-ionising region \citep[][]{Korista1997}. \citet[][]{Ferland2020} argue for the use of He~II~$\lambda$4686 in particular, as a relatively clean metric for the strength of the FUV continuum. Based on photo-ionisation modelling, \citet{Eastman1985} found that in BLR conditions, photons for He~II resonance lines, in particular He~II Ly$\alpha$, have a high probability of being absorbed by H~I. This means that, unlike the Balmer lines, which are affected by several complicating factors, He~II recombination is in good approximation to Case~B for Hydrogen\footnote{In Hydrogen Case B recombination, the ionised gas is assumed to be optically thin at photon energies below the Lyman limit.} \citep[][]{Ferland2020}. He~II~4686 therefore functions as an FUV photon counter: it provides our most direct measurement of the strength of the ionising continuum. The fact that in MKN~110 we observe the He~II flux to show {\em larger} variations than the optical continuum agrees with this idea. In the following we will therefore explore the use of the He~II $\lambda$4686 flux as a proxy for the FUV ionising continuum of MKN~110.

\begin{figure*}
\includegraphics[width=.75\textwidth]{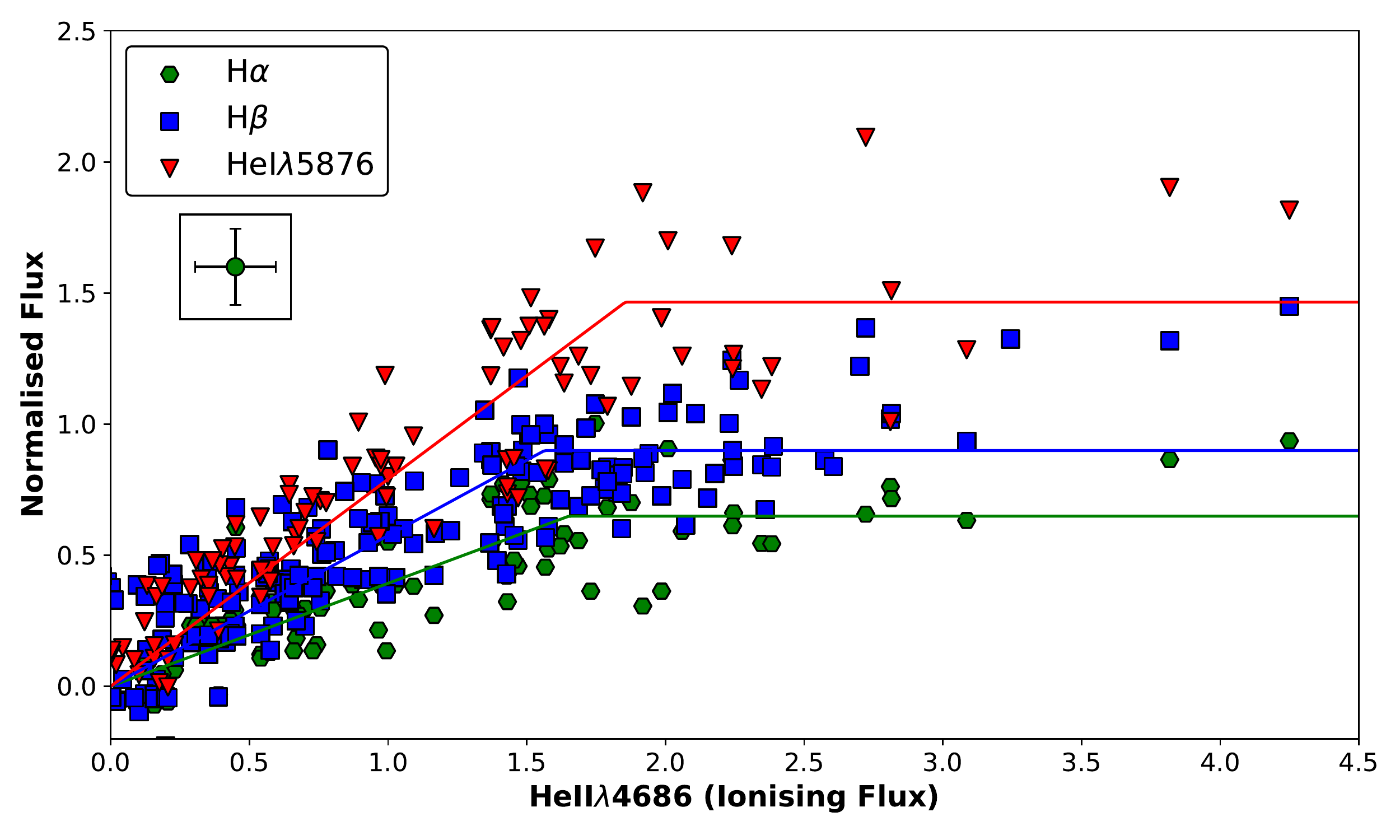}
\caption{The normalised line fluxes for H$\alpha$ (green hexagons), H$\beta$ (blue squares), and He~I $\lambda$5876 (red triangles), plotted against the normalised He~II $\lambda$4686 flux. All line fluxes have been corrected by adjusting for the non-variable narrow-line component (see Sections~\ref{sec:flux_corr} and~\ref{sec:response_nonres}). The responses of the line fluxes have been modelled using a two-component fitting function. The fitting results show the line species have different gradients in their response, as well as different saturation levels. This stratification indicates slope and saturation level are efficient parameters for distinguishing the three lines species. The uncertainties on the data were calculated by incorporating the intrinsic scatter apparent in the data (see Section~\ref{sec:response_quant}). We include the uncertainties on the data-set for H$\alpha$ as an example, in the top left of the plot.}
\label{fig:mrk110_responsefit}
\end{figure*}

\subsection{Correcting for the non-responsive flux component}
\label{sec:response_nonres}
Figure~\ref{fig:mrk110_responsefit} shows normalised H$\alpha$, H$\beta$, and He~I~$\lambda 5876$ plotted against He~II~$\lambda 4686$. The fluxes for the three lower ionisation lines have been corrected for the presence of a non-varying component in the following way. Each line species showed a vertical offset, indicating that at the lowest He~II~$\lambda 4686$ flux states, there was still a measurable line flux. This is despite the fact that at the lowest flux state (associated with the SDSS spectrum), the He~II~$\lambda$4686 flux had completely disappeared. We correct the fluxes by removing this non-variable component, as our interest is in the fraction of the flux that is responsive to He II changes. Unlike for He~II~$\lambda$4686, we have not subtracted a narrow line flux from the broad component for H$\alpha$, H$\beta$, and He~I~$\lambda 5876$ (Section~\ref{sec:flux_corr}). This means that the non-varying component in these lines must at least partially be associated with narrow line emission. However, it will also include any non-varying broad-line flux.

The correction for the non-varying flux contribution is made by performing a linear fit to the flux data for each line species in the range where normalised $f_{HeII}$ $< 1$. The offset of the data is such that it requires the linear fit to pass through the origin (see Figure~\ref{fig:mrk110_responsefit} for the effect in the corrected data-set). The downward shifts in units of normalised flux are 0.2, 0.35 and 0.45 for He~I $\lambda$5876, H$\beta$, and H$\alpha$ respectively.

\subsection{Quantifying the response}\label{sec:response_quant}
It is evident from Figure~\ref{fig:mrk110_responsefit} that, above a threshold, the fluxes of all three line species, H$\alpha$, H$\beta$, and He~I~$\lambda 5876$, have a diminishing response to an increase in He~II $\lambda$4686 flux. For the highest He~II fluxes the response appears almost flat. Each line species responds approximately linearly at low luminosities and saturates to a fixed maximum flux as the He~II line increases further. There are also notable differences among the line species: both the gradient of the response at low He~II fluxes and the saturation level of the fluxes increases from  H$\alpha$ and H$\beta$ to He~I. A diminishing line response was also noted for the C~IV~$\lambda 1550$~line compared to the 1350\AA\ continuum \citep[][]{Wamsteker1986}. And similar behaviour was described by \citet{Ferland2020} for the He~II $\lambda 4686$~line itself. Using a low-redshift AGN sample spanning a very large range in luminosity, \citet[][]{Ferland2020} showed that the highest-luminosity sources, identified mainly with NLSy1s, showed the largest deficit in He~II emission, relative to predictions based on the SED.  

To quantify the observed behaviour, we fit the flux data with a simple empirical function. 
We experimented with a number of functions
and found that a double straight line results in the the most stable fits whilst also capturing the two key aspects of the response: a difference in gradient at low He~II fluxes and a difference in saturation level. If we let $y$ represent the normalised line flux for each of the the low-ionisation lines and $x$ the line flux for normalised He~II $\lambda$4686, our fitting function has the form
\begin{equation*}
    y(x) =
    \begin{cases}
    ax &  \text{if }x\leq x_k\\
    y_{sat} & \text{if }x>x_k .
    \end{cases}
\end{equation*}
Here $x_k$ represents the `knee' value of the He~II flux, where $y(x_k) = y_{sat}$, i.e. the turning point in the response. The free parameters are the slope $a$ and the saturation level $y_{sat}$, with $x_k$ fixed by the other two parameters.

To find the uncertainties on our fits, we need to account for the uncertainties in our data-set. We note that the measurement uncertainties, associated with the statistical uncertainties of the spectral fits at each epoch, represent an underestimate of the real variance. We are interested in the gradual, long-term evolution of MKN~110, which means that the uncertainties of interest are a combination of the measurement uncertainties and the scatter around the long-term behaviour caused by short-term (days--months) changes in the line-forming regions. We find that this scatter, evident in Figure~\ref{fig:mrk110_responsefit}, is the dominant cause of uncertainty in our data-set. To estimate the level of uncertainty, we first fit our model function using the statistical errors of the data. We then apply the standard deviation of the residuals as a uniform error on the data-set (for each line species separately). We then use this new error estimate, which now includes the additional uncertainties associated with short-term variability, to calculate the best-fit parameters of our fitting model. The results of our fits, as well as the estimated 2$\sigma$ uncertainties, are listed in Table~\ref{table:resp_fit_results} and the best-fit models are included in Figure~\ref{fig:mrk110_responsefit}.

\begin{table}
    \renewcommand*{\arraystretch}{1.5}
    \centering
    \caption{Fitting results long-term BLR responsivity curves. 
    }
    \begin{tabular}{lcccc}
    \toprule
    Line & $\sigma_{est}$~$^a$ & $a$~$^b$ & $y_{sat}$~$^c$ & RM lag~$^d$ \\
    \midrule
    \midrule
    H$\alpha$ & 0.15 & 0.39 $^{+0.06}_{-0.04}$ & 0.65 $^{+0.09}_{-0.08}$ & 32.3 $^{+4.3}_{-4.9}$ \\
    H$\beta$ & 0.18 & 0.57 $^{+0.04}_{-0.06}$ & 0.90 $^{+0.10}_{-0.06}$ & 24.2 $^{+3.7}_{-3.3}$ \\
    He~I~$\lambda$5876 & 0.24 & 0.79 $^{+0.07}_{-0.07}$ & 1.47 $^{+0.16}_{-0.15}$ & 10.7 $^{+8.0}_{-6.0}$ \\
    \bottomrule
    \end{tabular}
    \flushleft{\scriptsize{
    a) Estimated uncertainty (normalised line flux).\\
    b) Best-fit value for the slope of the two-component function.\\
    c) Best-fit value for the saturation level of the two-component function.\\
    d) Reverberation Mapping lags presented in K01 (days).
    }}
    \label{table:resp_fit_results}
\end{table}

\begin{figure}
\centering
\includegraphics[scale=0.95]{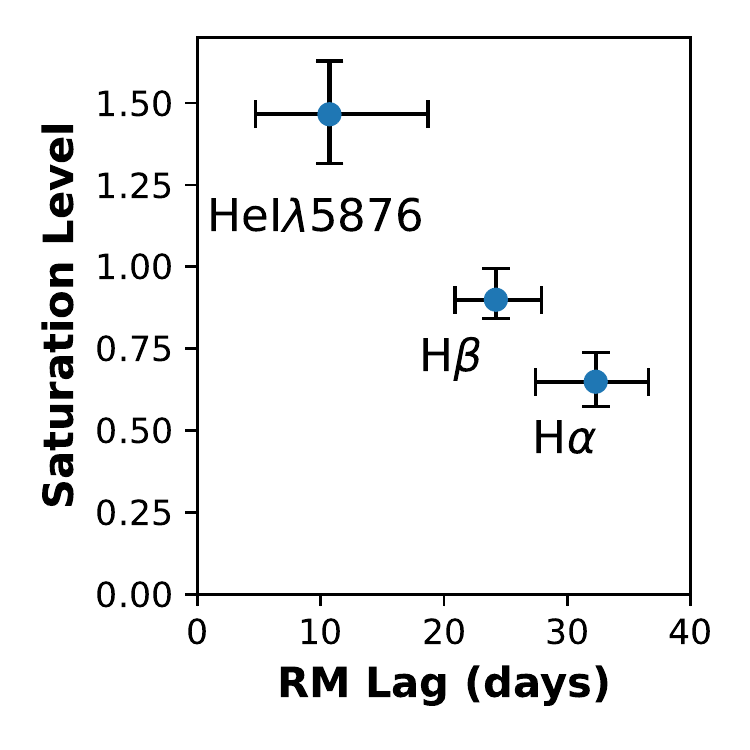}
\caption{The saturation levels ($y_{sat}$) for H$\alpha$, H$\beta$, and He~I $\lambda$5876, plotted against the Reverberation Mapping lags for these line species presented in K01. The saturation levels belong to the two-component model for the response of the normalised line fluxes to changes in the He~II~$\lambda$4686 flux (Section~\ref{sec:response_quant}). The results show a strong negative correlation between an emission line's ability to track the continuum and the distance of the line-forming region from MKN~110's central engine.}
\label{fig:mrk110_satvslag}
\end{figure}

The response results show an interesting anti-correlation when combined with the measured lags from the K01 RM study. These lags represent the approximate response time of each line species to changes in the FUV continuum and are therefore directly related to the distances between the SMBH and the different line forming regions. We plot our values of $y_{sat}$ against the RM lags in Figure~\ref{fig:mrk110_satvslag}. He~I~$\lambda$5876, which is formed closer to the central engine, saturates at a higher level than the other two lines. Similarly H$\beta$ saturates at a higher level than H$\alpha$. The ability of a line species to track the increasing FUV continuum therefore appears closely related to the physical parameters that determine the optimal line-forming distance from the SMBH. This conclusion can be drawn without making any assumptions about the nature of these physical parameters: we only need to assume that the RM lag corresponds to the distance from the central source of ionising radiation.

\subsection{Responsivity from epoch to epoch}
\label{sec:response_hysteresis}
To estimate the uncertainties on our line-responsivity fits (Figure~\ref{fig:mrk110_responsefit}) we made the assumption that a combination of short-term variability and measurement uncertainties led to a random scatter around a single long-term response curve. This approach allows us to treat all the spectroscopic data, which cover over thirty years, as a single data-set. Given the time-span of the combined data-set, we now investigate whether a trend is visible on timescales of several years. Over the three decades of observations, MKN~110 varies from high to low luminosity states. When MKN~110 returns to a luminosity state that it was previously in, do the line fluxes return to the same level they had before? If not, is this a form of hysteresis, in which the AGN `remembers' its evolution over several years? To answer these questions, we consider the difference in responsivity among different observational epochs. 
We define three such periods (see Figure~\ref{fig:lcf5100}): 
\begin{enumerate}[label=\Roman*)]
    \item MJD 46900--50900, covering BK99 and WHP98.
    \item MJD 51500--52300, covering K01, FAST-RM, and SDSS.
    \item MJD 53700-58500, new FAST and WHT data.
\end{enumerate}
In the first and last epochs, MKN~110 is on average in a higher flux state than during the middle epoch, in which MKN~110 dims to the overall lowest state (SDSS). Although the cadence of the spectroscopic coverage in epochs I and III is not as high as in epoch II, it is unlikely that all of our observations will have missed historic low states. To explore whether the long-term evolution of the luminosity state, from epoch to epoch, leaves a detectable imprint on the BLR and the line responsivities, we consider the response curves for each of the line species separately.

\begin{figure*}
\centering
\includegraphics[width=.7\textwidth]{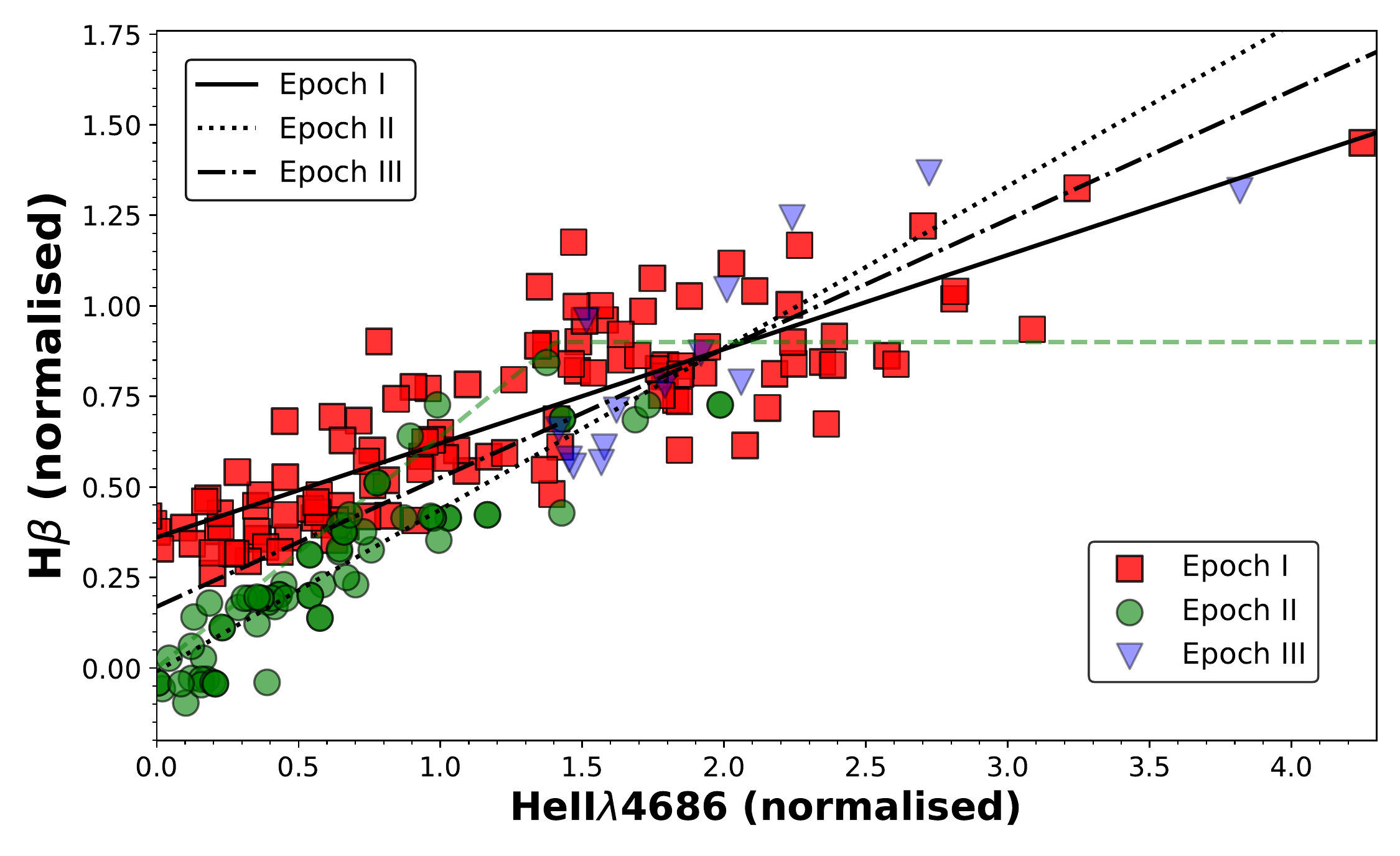}
\caption{The normalised H$\beta$ flux data plotted against He~II$~\lambda$4686, split according to epoch
as in Figure~\ref{fig:lcf5100}. The dashed green line is our best-fit two-component model for the full H$\beta$ data-set as seen in Figure~\ref{fig:mrk110_responsefit}. The three straight lines represent the results of linear regressions on subsets of the data: continuous, dotted and dot-dashed lines for epochs I, II, and III respectively. It is noticeable that the slope in the data points for the second epoch, associated with a relatively low state, is different from those of the two `high state' epochs. This implies that identical changes in He~II flux resulted in different changes in H$\beta$ flux, with a time interval of 5--10 years, from one epoch to the next (see Figure~\ref{fig:lcf5100}). On shorter timescales, within a single epoch, this effect is not visible (Section~\ref{sec:response_hysteresis}). The slopes of the linear regressions indicate that there is also a difference in response between the two high-state epochs.}
\label{fig:mrk110_origin_fit}
\end{figure*}

In Figure~\ref{fig:mrk110_origin_fit} we show the example of H$\beta$ (plotted against He~II as in Figure~\ref{fig:mrk110_responsefit}), where the data have now been grouped by epoch.
He~I~$\lambda$5876 and H$\alpha$ show similar patterns as H$\beta$. The data from epoch I show a shallower response than those of epoch II. As the data in epoch II and III do not cover a sufficiently broad range in normalised He~II flux, we compare the responsivity among epochs using a simple linear regression. The results of the linear fits for the three line species are listed in Table~\ref{table:epoch_fit}. The fitting results confirm the change in response over time: we measure a significantly steeper responsivity for epoch II as well as a difference in slope between epoch I and III. The stronger response for epoch II is as expected, as this epoch is dominated by low flux-state data points (the sloped part in our two-component model). The difference in slope between epoch I and epoch III suggests that the responsivity does not only evolve with flux state, but also over timescales of 5--10 years.

The data-sets comprising epoch II provide our most detailed view of the evolution of continuum and BLR in MKN~110. We examined whether there is any significant difference among the different seasons within this middle epoch, which cover shorter periods (up to 1 year) of rising or lowering continuum levels. Linear regressions on subsets of the data show that the slopes are all consistent within uncertainties. 
It therefore appears that the BLR responsivity changes only on longer timescales. In general, responsivity of the BLR to changes in ionising continuum will be due to either lighting up different parts of a static but distributed structure, or to the BLR physically changing. The fact that we see repeatability on short timescales but differences on long timescales suggests that in fact {\em both} of these effects are happening. We will return to this point in Sections~\ref{sec:discussion_timescales}~and~\ref{sec:discussion_patterns}.

\begin{table*}
    \renewcommand*{\arraystretch}{1.5}
    \centering
    \caption{Results of a linear regression of the normalised line flux data, separated by line species and by epoch.
    }
    \begin{tabular}{lcccccccccc}
        \toprule
         & Full & \multicolumn{3}{c}{I} &\multicolumn{3}{c}{II} & \multicolumn{3}{c}{III} \\
        Line & Shift~$^a$ & Slope & Offset & R$^2$~$^b$ & Slope & Offset & R$^2$ & Slope & Offset & R$^2$ \\
        \midrule
        \midrule
        H$\alpha$ & 0.45 & 0.10$\pm$0.04 & 0.46$\pm$0.08 & 0.26 & 0.27$\pm$0.03 & 0.04$\pm$0.02 & 0.60 & 0.13$\pm$0.07 & 0.37$\pm$0.14 & 0.23 \\
        H$\beta$ & 0.35 & 0.26$\pm$0.02 & 0.36$\pm$0.02 & 0.71 & 0.45$\pm$0.02 & $-$0.01$\pm$0.02 & 0.83 & 0.36$\pm$0.07 & 0.17$\pm$0.15 & 0.67 \\
        He~I~$\lambda$5876 & 0.20 & 0.26$\pm$0.06 & 0.69$\pm$0.12 & 0.48 & 0.65$\pm$0.04 & 0.13$\pm$0.03 & 0.80 & 0.48$\pm$0.14 & 0.40$\pm$0.29 & 0.49 \\
        \bottomrule
    \end{tabular}
    \flushleft{\scriptsize{
    a) Shifts applied to the normalised line fluxes (Section~\ref{sec:response_nonres}).\\
    b) Coefficient of determination (1--$\sum (residuals)^2$/Variance).
    }}
    \label{table:epoch_fit}
\end{table*}


\section{Line profiles and kinematic evolution}
\label{sec:offsets}


The broad components of the emission lines in MKN~110 are redshifted by velocities of 100s to 1000s~km~s$^{-1}$, relative to their narrow-line components and therefore relative to the redshift of the AGN and its host. This broad-to-narrow-line offset is most distinct for He~II~$\lambda$4686, but is also visible as a line asymmetry in the Balmer lines (Figure~\ref{fig:fit_example}). As noted in Section~\ref{sec:mkn110_history}, the offset has been associated with a gravitational redshift (K03b). However, the large range among different mass estimates for the SMBH in MKN~110 leads us to investigate alternatives. We first consider whether the offset in the emission lines, like the line flux, depends on the level of the FUV ionising continuum. We focus on two lines: He~II~$\lambda$4686 and H$\beta$. As we are investigating changes in the offset over time, we follow the fitting procedure set out in Section~\ref{sec:fitting_profiles} (in contrast to the procedure in K03b, who use the MAD spectrum). The large amount of available spectra in our data-set allows us to track the line evolution over time.

\subsection{Offset, FUV flux, and line width}

The data-set used for this part of the study contains all FAST and WHT spectra. As discussed in Section~\ref{sec:fitting}, the lowest state spectra are not included individually. These spectra are included as an average, established by stacking the spectra, such that they have sufficient S/N for spectral fitting. All spectra in FAST-RM1 as well as the first spectrum of FAST-RM2 (MJD 52940) are stacked together. The other spectra are fit individually. Before comparing the narrow line to broad line offset, we confirm that the narrow line components of He~II and H$\beta$ are found at the same redshift. For He~II, the centre of the narrow line was kept fixed in the fitting procedure (Section~\ref{sec:fitting_profiles}), at a value corresponding to $z=0.0351$. For H$\beta$, we find that the median value of the narrow line centre, over all fitted spectra, also corresponds to $z=0.0351$. The standard deviation of the H$\beta$ narrow-line centres is $\sim$0.5{\AA}, corresponding to around 28~km~s$^{-1}$. This difference is very small compared to the measured offsets between narrow and broad components. In addition to the similarity in redshifts, we find that the width ($\sigma$) of the narrow-line components is comparable within 10\%: 4.2{\AA} for He~II and 3.8{\AA} for H$\beta$. We therefore conclude that the narrow-line components for He~II and H$\beta$ are formed in the same part of the Narrow Line Region and can be reliably used to gauge the offsets of the broad line components. 

\begin{figure*}
\centering
\includegraphics[width=\textwidth]{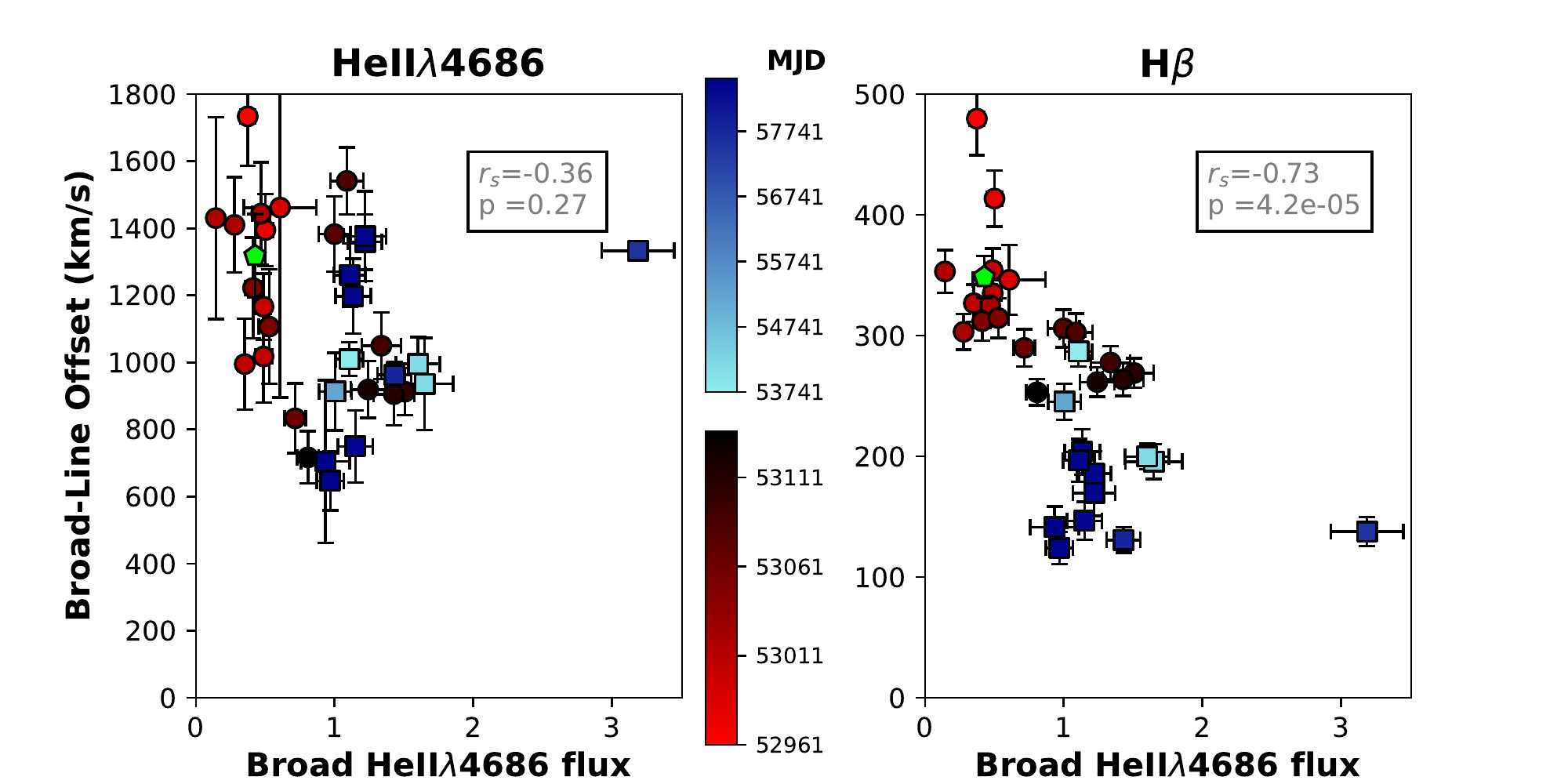}
\caption{The offset between the line centres of the broad and narrow components of He~II $\lambda 4686$ (\emph{left}) and H$\beta$ (\emph{right}), plotted against the normalised broad He~II flux. The offset between between the broad and narrow lines measures the relative redshift of the broad line components. The broad He~II flux acts as a proxy for the ionising continuum. The shading of the colours indicates the time at which the spectrum for the data points was taken. The two colour ranges represent the observational periods: red circles cover the spectra of the FAST-RM2 campaign (MJD 52961--53137) and blue squares cover the newer FAST and WHT spectra (MJD 53741--58500). The green hexagon represents the fits from the stacked `low-state' FAST spectra (MJD 52252-52798). $r_s$ is the Spearman ranked correlation coefficient. The $p$ value is for a two-tailed test (on a null hypothesis of non-correlation) of Pearson's correlation coefficient and it indicates the probability that the two parameters are not correlated. Both $r_s$ and the $p$-value relate to the the full data-sets. The He~II data show two distinct `branches' as the ionising flux changes over time, whereas H$\beta$ shows a consistent correlation.}
\label{fig:mrk110_flux_offset}
\end{figure*}

The results of our spectral fitting indicate that the velocity redshift of broad-line components compared to the narrow lines changes significantly over time, for both He~II and H$\beta$. The correlations among offset and He~II/FUV flux present a complex picture of MKN~110's broad-line variability. Figure~\ref{fig:mrk110_flux_offset} shows the velocity offset between components plotted against broad He~II flux. The left panel shows the results for He~II and the right panel the results for H$\beta$. The data have been split between the FAST-RM2 data (red circles) and the newer FAST and WHT data (blue squares). The epoch is indicated by the shading of the colours, going from light to dark for later epochs. In the H$\beta$ plot, one can trace the passage of time from the top left (light red circles) to the bottom right (dark blue squares). The observed increase in He~II flux agrees with the rise in the continuum lightcurve for this period (See epochs II and III in Figure~\ref{fig:lcf5100}).

The first thing to note in Figure~\ref{fig:mrk110_flux_offset} is that the average offset for He~II is significantly larger than for H$\beta$. Second, H$\beta$ appears to have a negative correlation between He~II flux and offset: as the He~II line flux increases, the offset becomes smaller. In other words: as the ionising flux goes up, the relative redshift of the broad component goes down. The correlation for H$\beta$ is consistent over the different epochs, whereas the He~II behaviour is less straightforward, as it appears to form two distinct branches. 
Finally, we note that not just the {\em average} offset, but the {\em change} in offset is larger for He~II than for H$\beta$. The somewhat aberrant high He~II flux point in the far right of both plots is associated with the WHT spectrum of MJD 57539, the spectrum with the highest S/N in the sample.

\begin{figure*}
\includegraphics[width=\textwidth]{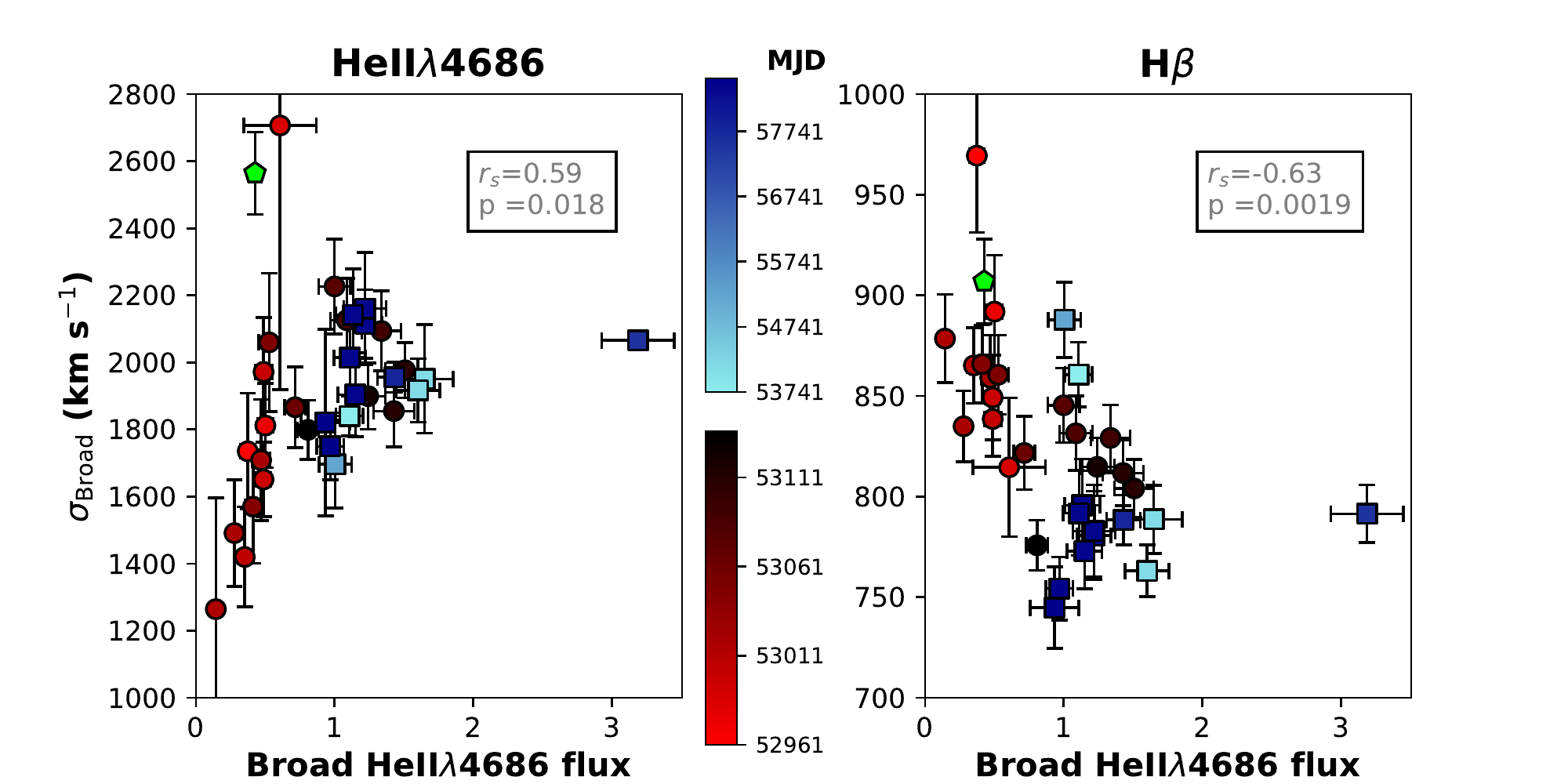}
\caption{The width of the broad emission-line components plotted against the broad He~II $\lambda$4686 flux. The colours, shading, and correlation metrics are the same as in Figure~\ref{fig:mrk110_flux_offset}. For H$\beta$ we can see that the broad-line width decreases with increasing He~II flux, an inverse correlation known as AGN breathing: as the luminosity (and the FUV flux) increases, the radius of the BLR increases, corresponding to a smaller rotational velocity and $\sigma_{broad}$ of the BLR clouds. Interestingly, He~II shows an opposite trend, which could indicate that the kinematics of the He~II line-forming region are not purely Keplerian.}
\label{fig:mrk110_flux_sigma}
\end{figure*}

We next consider the correlation between the broad-line widths and the He~II $\lambda$4686 flux. The results are shown in Figure~\ref{fig:mrk110_flux_sigma}. The line widths for He~II are significantly larger than for H$\beta$, in agreement with the notion that He~II is formed closer to the ionising source (as found in RM studies). A luminosity-dependent line width, as would be the case for a `breathing' BLR, is evident for H$\beta$. In agreement with the results shown in Figure~\ref{fig:mrk110_flux_offset}, the distance of the broad-H$\beta$ emitting region to the SMBH (which is inversely correlated with the broad-line width) increases with luminosity. This breathing of the BLR is expected, if the BLR motion is Keplerian. This comparison functions as a check on both our line-fitting results and the basic assumptions made in the analysis presented so far. Remarkably, He~II shows the opposite trend. As was the case in the offset-vs-flux results, the data for H$\beta$ form a clearer pattern than those for He~II.

\begin{figure*}
\includegraphics[width=\textwidth]{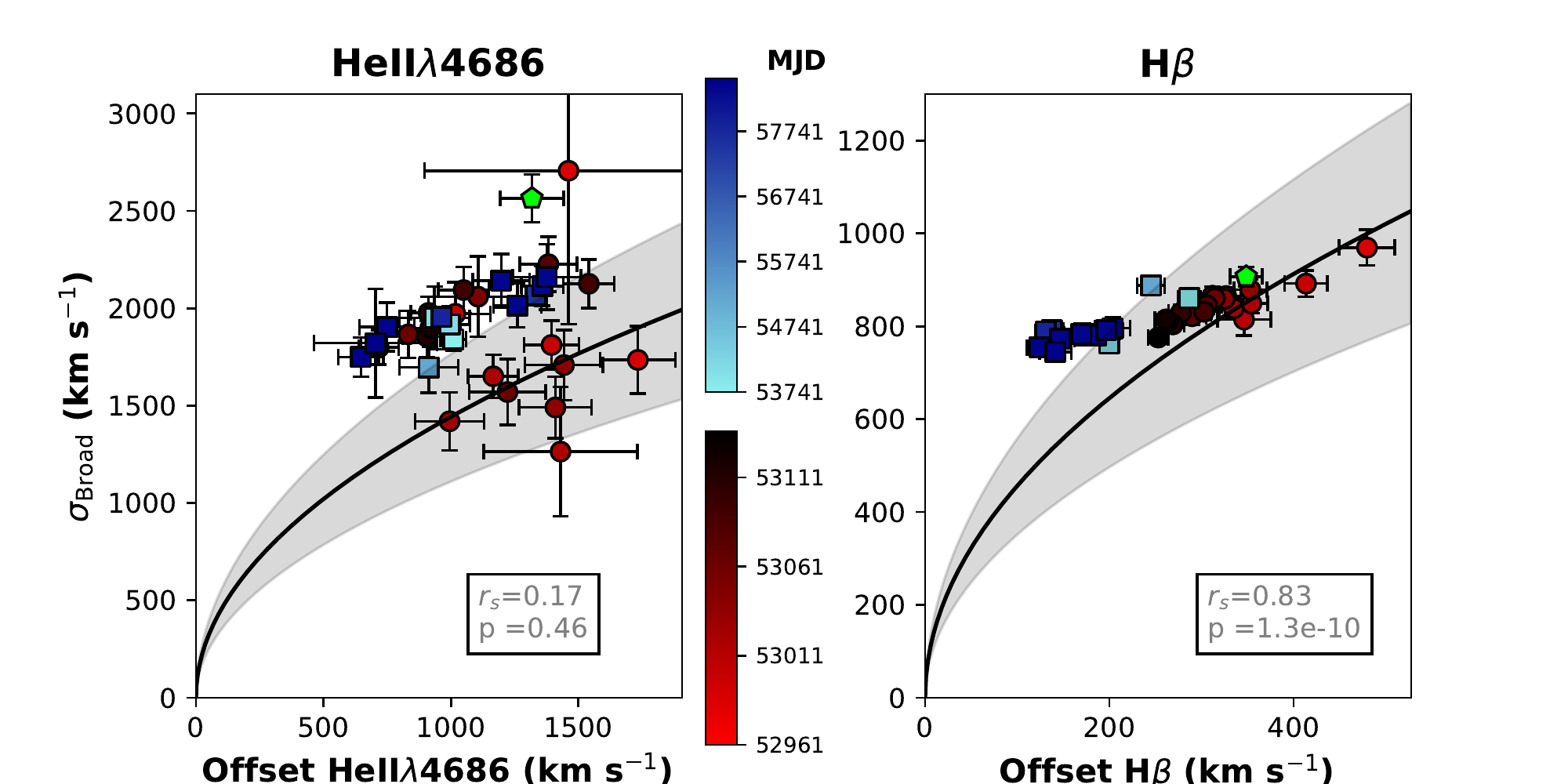}
\caption{The line width of the broad component of He~II $\lambda 4686$ (\emph{left}) and H$\beta$ (\emph{right}), plotted against the broad-to-narrow-line offsets. The colours, shading, and correlation metrics are as described for Figure~\ref{fig:mrk110_flux_offset}. There is a strong correlation between the parameters for H$\beta$, whereas the pattern for He~II is considerably noisier. The black lines indicate the theoretical relation between offset and $\sigma$, if the offset is caused by gravitational redshift, for the case of a 21$^\circ$ inclination and virial factor $f=4.1$. The shaded regions correspond to $\pm 5^\circ$ uncertainty, following the estimate from K03b.}
\label{fig:mrk110_offset_sigma}
\end{figure*}

Finally, we investigate the dependence of the broad line widths on the offsets between the line components. The results are shown in Figure~\ref{fig:mrk110_offset_sigma}. We find that there is a clear correlation between the two parameters for H$\beta$, whereas there is again 
no single pattern in the He~II data. If the offset is inversely correlated with the distance from the line-forming region to the SMBH, the behaviour of H$\beta$ would be a typical presentation of AGN breathing. He~II remarkably shows a trend in the opposite direction. This could indicate that the motion of the BLR where He~II is formed is not purely set by Keplerian 
kinematics. We return to the discussion and interpretation of this behaviour in Section~\ref{sec:discussion_kinematic_response}.


\subsection{Origin of scatter}
The data in Figures~\ref{fig:mrk110_flux_offset}, \ref{fig:mrk110_flux_sigma}, and~\ref{fig:mrk110_offset_sigma} show considerable scatter. 
The data for H$\beta$ appear to show a clearer pattern than those for He~II. A potential cause for the scatter that is extrinsic to MKN~110 is the relative quality of the fits: the He~II line is considerably weaker than H$\beta$ in all spectra, making the fit more sensitive to noise. This interpretation agrees with the fact that the scatter in Figure~\ref{fig:mrk110_flux_offset} appears largest for the spectra at earlier epochs, when the He~II line was relatively weak.

However, the scatter does not appear to be random: whereas the stacked result, which is made up almost exclusively of FAST-RM1 data, is consistent with the trend of the FAST-RM2 data, the later high-state data seem to be on a separate `branch'. This separation is much more significant for He~II than for H$\beta$. If the scatter in the He~II data is intrinsic to the AGN, this could imply that the response to a changing continuum is more erratic for gas closer to the central engine (such as for the gas that forms He~II compared to the gas forming H$\beta$), or that the response of this gas can change on shorter timescales. We return to this point in Section~\ref{sec:discussion_timescales}.

\section{Summary of key results}
\label{sec:keyresult}

\subsection{Broad-Line Flux Response effects}
\label{sec:keyresult-flux}


\textbf{R1.1: Large variability}. MKN~110 shows strong variability on timescales of months to decades, in both the continuum (a factor of $\sim$10) and the lines (up to a factor $\sim$40 for He~II~4686).\\

\noindent \textbf{R1.2: Extreme He~II variability}. Among the emission lines in the optical range, the variability is most pronounced for the high-ionisation line He~II $\lambda$4686. This is in agreement with previous AGN variability studies, which show that the amplitude of variability correlates with the ionisation level \citep{Peterson2008,Bentz2010}. The He~II variability is \textit{considerably stronger}, around a factor of 4, than that of the optical continuum, suggesting it is a proxy for the inaccessible FUV continuum variability.\\

\noindent\textbf{R1.3: Differentiation in the emission line response}. As the He~II flux heads towards zero, other lines do not, with shifts of the order of 20-40\% of the normalised line flux. We correct for this non-responsive component by subtracting the fitted He~II~4686 narrow-line flux (Section~\ref{sec:flux_corr}) and an offset for H$\alpha$, H$\beta$, and He~I~$\lambda$5876 (Section~\ref{sec:response_nonres}).\\

\noindent\textbf{R1.4: Curvature in the emission line response}. The response of the H$\alpha$, H$\beta$, and He~I $\lambda$5876 emission lines to changes in the FUV flux (as indicated by He~II $\lambda$4686) is dependent on the level of incident flux. The line-emitting gas is less responsive at high FUV flux levels (Figure~\ref{fig:mrk110_responsefit}). The effect can be approximated by a two-component function that has a flat response once it reaches a saturation level (Figure~\ref{fig:mrk110_responsefit}). It is important to note that the full data-set covers $\sim$30 years of observations. This means that the responsivity discussed here is different from the instantaneous responsivity previously investigated in photoionisation models \citep[e.g.][]{OBrien1995,Korista2004}.\\

\noindent\textbf{R1.5: Saturation-distance correlation}. The level at which the response of the emission lines saturates, shows a negative correlation with the lags found in RM measurements (Figure~\ref{fig:mrk110_satvslag}). This suggests that the saturation level for the lines is related to the physical location in the BLR or that it depends on the minimum ionisation energy required to generate the line.\\

\noindent\textbf{R1.6: Historical changes}. 
The response of the lines to changes in the FUV differs from epoch to epoch. Over a timescale of less than 5 years, the response curve of a line is repeatable (i.e. going up and down in FUV flux has the same results), whereas data separated by 5--10 years show a different line response (Figure~\ref{fig:mrk110_origin_fit} and Table~\ref{table:epoch_fit}). A notable aspect of this slow-moving evolution is that some of response curves in Figure~\ref{fig:mrk110_origin_fit} do not return to zero, even after our earlier removal of an average offset. This means that the fluxes from both the responsive and non-responsive material change with historical epoch.



\subsection{Kinematic effects}
\label{sec:keyresult-kinematic}

\textbf{R2.1: Velocity widths and velocity offsets change}. As noted by K03b, the line profiles of He~II~$\lambda$4686 and H$\beta$ show a clear offset between the broad and narrow components. K03b, using a data-set covering $\sim$7 months, found that the offset is smaller for lines which have longer reverberation lags. In this study, where we make use of spectra taken over a period of $\sim$16 years, we find that the magnitude of both the offsets and the line widths evolve over time (Figures~\ref{fig:mrk110_flux_offset}~and~\ref{fig:mrk110_flux_sigma}).

\noindent\textbf{R2.2: Velocity offset vs. flux}. The broad-to-narrow-line velocity offset shows a negative correlation with the broad He~II flux for the H$\beta$ line (Figure~\ref{fig:mrk110_flux_offset}). This means that the velocity offsets are smaller at high flux states. The correlation is visible for H$\beta$, whereas the response of He~II is more complex and possibly double-branched.

\noindent\textbf{R2.3 Line width vs. flux}. The line width shows a negative correlation with the He~II flux for H$\beta$ (Figure~\ref{fig:mrk110_flux_sigma}).  He~II, in contrast, shows overall a positive correlation, and again the behaviour is more complex.

\noindent\textbf{R2.4: Line width vs. offset}. The line width of the broad component shows a strong and clear positive correlation with the offset for H$\beta$ (Figure~\ref{fig:mrk110_offset_sigma}). For He~II we find no significant correlation. 

\noindent\textbf{R2.5: Contrast between He~II and H$\beta$}. As the ionising flux varies, the H$\beta$ line shows quite simple and consistent changes in velocity offset and line width. He~II on the other hand shows no such clear trends. Although this can reflect uncertainties in the fitting process, this clear distinction likely indicates the large differences in kinematics between the two line-forming regions. 

\section{Discussion: General Points}
\label{sec:discussion-general}

In the following discussion sections, we concentrate on 
model-independent implications, with some comments on specific models.

\subsection{Black hole mass and scale alternatives}
Interpretations depend on the observed lags and deduced masses, so we first summarise these. Based on the lags observed by K01, the BLR is clearly stratified, with the material dominating the He~II, H$\beta$, and H$\alpha$ emission located at distances of 3.9 $^{+2.8}_{-0.7}$, 24.2 $^{+3.7}_{-3.3}$, and  32.3 $^{+4.3}_{-4.9}$ light-days respectively. However, the physical implications of these lags also depend on the assumed black hole mass. As discussed in section~\ref{sec:mkn110_history}, K02 and K03b represent the two extreme ends of the distribution of mass estimates.
Based on the lags and the assumption that the line widths are set purely by Keplerian motion, KB02 derived $M_1=1.8\times 10^7 M_\odot$. The above lags then correspond to $R/R_S= 1,900, 11,800$, and $15,700$ for He~II, H$\beta$ and H$\alpha$. Here $R_S$ is the Schwarzschild radius. On the other hand, based on the assumption that line velocity offsets are caused by gravitational redshift, K03b derived $M_2=1.4\times 10^8 M_\odot$. The line-lags then correspond to $R/R_S= 244, 1,516$, and $2,023$. To make this second mass value consistent with the observed lags, K03b discusses a disc-wind model, in which we are seeing the MKN~110 system almost pole-on, at an inclination of 21$^\circ$. In this case, there could well be a significant additional geometric correction factor in deducing the distance of the BLR clouds from the observed time lags. This large correction factor would mean the black hole mass deduced from the lags would also be significantly larger, reducing the tension between the gravitational-redshift mass estimate and the lag-based mass estimate.

\subsection{Is MKN~110 representative?}
\label{sec:discussion_representative}
There are two ways in which MKN~110 may be considered unusual for a broad-line type 1 AGN - the first is its large amplitude of variability and the second is the narrowness of the broad lines. Regarding the variability, there are other well-known historic examples of large changes in individual sources, for example NGC~1566 \citep{Alloin1985}, NGC~3516 \citep{Ilic2020}, Cen~A \citep{Lawrence1977}, and NGC~4151 \citep{Penston1984}. Recently, it has become clear that even high-luminosity quasars can show extreme variability over decadal timescales. In a sample of approximately 100,000 quasars, \citet[][]{MacLeod2016} found that $\sim$1\% show magnitude changes $|\Delta g|>1$ on timescales smaller than 15 years. Similarly, an estimated fraction of 30--50\% of quasars shows variability of $|\Delta g|>1$ on timescales greater than 15 years \citep[][]{Rumbaugh2018}, suggesting large changes in AGN are perhaps quite common.

Regarding line width, the Balmer lines are certainly narrower than for most type 1 AGN, and towards the upper end of what would traditionally be classed as NLSy1 \citep[][]{Osterbrock1985,Komossa2008}, although the He~II width is more typical of AGN BLRs.
For MKN~110 FWHM$_{H\beta}\sim$1940 km s$^{-1}$, on average; for NLSy1 typically FWHM$_{H\beta}<2000$~km~s$^{-1}$ \citep[][]{Osterbrock1985}, whereas `regular' type 1 Seyferts have broad-line widths of thousands to tens of thousands km s$^{-1}$ \citep[][]{Peterson_BLR-review-2006}. Traditionally, there have been two rival explanations of the NLSy1 phenomenon: normal AGN seen close to pole-on, or objects with a high Eddington ratio, such that lines are formed further out. The current consensus is that the high-Eddington explanation is broadly correct, with perhaps a minority of objects being pole-on cases \citep[][]{Komossa2008}. MKN~110 could fit into either explanation. If the larger mass $M_2$ is correct, then we are seeing  MKN~110 at an inclination angle of 21$^\circ$. An inclination this small or smaller will occur in $\sim$6\% of all AGN, for random orientations. Alternatively, if the smaller mass $M_1$ is correct, then MKN~110 is marginally super-Eddington - $L/L_E\sim 1.4$ using the optical luminosity estimated by \citet{Bischoff1999} and a standard bolometric correction \citep[][]{Richards2006}. As noted in Section~\ref{sec:mkn110_history}, MKN~110 has very weak Fe~II emission, and a normal X-ray spectrum and X-ray loudness, whereas most NLSy1s have strong Fe~II, steep X-ray spectra, and are X-ray quiet \citep[e.g.,][]{Lawrence1997}. This may suggest that the ``pole-on'' explanation is more likely; however, it is not conclusive.

In either of the above scenarios, MKN~110 is indeed unusual, but it remains a useful laboratory, to test ideas about the structure and physical nature of the BLR.

\section{Discussion: Response timescales in the BLR}
\label{sec:discussion_timescales}

We observe non-linear line response repeatable over short timescales, but not over longer ones (results R1.4, R1.5, R1.6).
To understand this, it is  useful to separate three kinds of BLR response - \textit{reverberation}, \textit{breathing}, and longer term \textit{structural response}. 

\subsection{Reverberation} 
At the high densities we understand to be present in BLR clouds, the recombination timescale is of the order of hours, so, locally, clouds track changes in the ionising continuum with negligible delay. We see those changes with a lag of the order of tens of days due to the finite size of the overall BLR system. The light curve of each line is a convolution of the continuum light curve, encoding the geometry and radial structure of the BLR. Using reverberation to deduce that structure is the focus of considerable effort in AGN monitoring. 

\subsection{Breathing of the BLR} 
For some AGN, it has been demonstrated that the observed lag changes from one reverberation campaign to another, and that the observed lag correlates with the mean luminosity during the campaign \citep{Peterson1998,Peterson2002}. The change in lag does not imply physical bulk motion; rather, this phenomenon is consistent with the basic insight of the `Locally Optimised Cloud' (LOC) picture of \citet{Baldwin1995}, that the BLR contains material with a wide range of distances, densities and thicknesses, with the observed emission representing a weighted mean response over all this material. Because the response is sensitive to the ionisation parameter, the radial location of that mean changes with luminosity. This phenomenon, of epoch-by-epoch shifting of the BLR response, has been referred to as `breathing' \citep[][]{Korista2004,Cackett2006}. The term `breathing' is somewhat misleading as it implies a bulk motion, rather than the actual interpretation of `lighting up' different regions within a fixed structure.

\subsection{Structural response} 
In the breathing phenomenon, a fixed BLR structure responds passively to the changing luminosity. We then expect that the line-flux response to changing luminosity should be repeatable from one season to another. This is indeed what we see on $\sim$year timescales, but on $\sim$decade timescales, the response pattern changes. This suggests that on long enough timescales, the BLR clouds themselves, and their distribution in space, physically respond to the radiation from the central engine. This raises the question whether this is consistent with physical expectations.

For a variety of models, restructuring may occur on roughly the \textit{dynamical timescale} $t_d=\sqrt{R^3/GM}$ - for example if BLR clouds are condensations formed in a disc wind driven by the central radiation \citep[][]{Murray1995,Elvis2000,Proga2000,Waters2016,Matthews2020}. In this case $t_d$ represents the time that material takes to travel through the BLR. At the H$\beta$-lag radius of 24.2 light days, for the smaller black hole mass case $M_1$ we get $t_d$=10.2~yrs, and for the larger mass $M_2$ we get $t_d$=3.6~yrs. Therefore it is not sufficient to simply observe over several years. What is required is to observe changes in the luminosity smoothed over several years. From Figure~\ref{fig:lcf5100}, it is clear that the long-term average has indeed changed from epochs I to II to III. We can then plausibly make the argument that we have detected the BLR dynamical timescale in MKN~110.

Another possible cause for the observed long-term evolution is that individual BLR clouds are compressed by the central engine radiation, and so adjust their density structure, such as in the model presented by \citet{Baskin-RPC-P4-2014} and \citet{Baskin2018}. This should occur on 
the \textit{acoustic crossing timescale} for a BLR cloud. For a cloud of thickness $N_H$, number density $n_e$, and temperature $T$ we get

$$t_a=2.7 {\rm yrs} \; 
\frac{N_H}{10^{23} {\rm cm}^{-2}} \;
\frac{10^9 {\rm cm}^{-3}}{n_e}.
$$ 

\noindent We therefore find that this is also a  potential explanation of changes in the long-term response pattern. The clouds in the \citet{Baskin-RPC-P4-2014} model will maintain $U$$\sim$1, as is expected for BLR material, even after significant changes in internal pressure.

A further clue comes from the kinematic changes seen in MKN~110's broad emission lines. Correlations are seen among line width, offset, and flux (results R2.2, R2.3, R2.4), but while the changes in $H\beta$ are simple and consistent, He~II shows a more complex behaviour (R2.5). This may be due to the shorter structural response timescale of the He~II emitting gas, causing hysteresis effects. For mass $M_1$, while the dynamical timescale at the H$\beta$ lag radius is 10.2~yrs, for the He~II lag radius of 3.9 light days we find a dynamical timescale of 240 days, comparable to an observing season, so that the structure of the He~II line-emitting regions may be substantially different from one year to the next, and even change significantly during a single season. 

Like the dynamical timescale, the acoustic timescale may also be shorter for clouds closer in to the ionising source (such as He~II forming clouds compared to H$\beta$ forming clouds). This would be the case if the typical density of BLR clouds increases towards the centre of the system. However, whether this effect occurs is less clear than the difference in kinematic timescales. Overall, the kinematic effects that we see thus add weight to the suggestion that in addition to the normal breathing, we are seeing the BLR physically respond on a mechanical timescale, i.e. corresponding to gross motions or pressure waves.

\section{Discussion: The shape of flux response patterns in the BLR}
\label{sec:discussion_patterns}

Grouping all epochs together, it seems that the BELs in MKN~110 have a curved response to changes in ionising luminosity, in that the emission lines are less responsive at high luminosity (result R1.4). However, even with our large and long-timescale data-set, it is still difficult to reliably distinguish instantaneous flux changes from historical changes (result R1.6). Nevertheless it seems clear that responsivity is slower than proportional, and varies between line species, even for recombination lines. Furthermore, the relation between line flux and ionising flux has a non-zero intercept, i.e. it seems that some of the BLR material does not respond to ionising flux changes at all, even at low luminosity. We note that Figure~\ref{fig:mrk110_responsefit} shows responses after correcting for the non-responsive component average over all epochs, and Figure~\ref{fig:mrk110_origin_fit} shows an epoch-dependent offset with respect to that global average offset.

\subsection{Single cloud response}
For the simplest single cloud models, thick clouds will be ionisation bounded, with an ionised zone on the side facing the continuum source. The emitted flux for Hydrogen and Helium recombination lines should then be roughly proportional to continuum luminosity as the ionised zone increases in depth with higher luminosities. For heavier elements, and collisional lines, things are more complicated. Even for Hydrogen, for high density clouds, and with an SED extending into the X-rays, there are effects that have plagued AGN modellers for years, such as ionisation from above the ground level, level-dependent line-transfer effects, and the effect of a partially ionised zone caused by X-rays \citep[see e.g.][]{Netzer-book-2013}. Nonetheless, line-luminosity should continue to increase with luminosity. 

Thinner clouds could be matter bounded, i.e. fully ionised, so that a change in luminosity may produce no change in line flux at all. To produce a curved response including saturation, a cloud of intermediate thickness could be ionisation bounded at low luminosity, and matter bounded at high luminosity, i.e. at high enough luminosity the ionising radiation can penetrate the far side of the cloud. This transition mechanism was also proposed by \citet[][]{Wamsteker1986} to explain the diminishing response of the C~IV~$\lambda 1550$~line to continuum increases. As pointed out by \cite{Gaskell2009}, rather than a single thick cloud, there could be a series of overlapping, and so self-shielding, thinner clouds at different distances. As the luminosity increases, the clouds become fully ionised in radial sequence, inner clouds first. 

\subsection{Distributed cloud systems}
As recognised in the LOC picture, the BLR is much more likely to be a system of clouds, with a wide distribution of densities, cloud thicknesses, radial distances, and covering factors. To calculate the net output requires photo-ionisation calculations over a grid of cloud properties.  In a series of papers, Goad, Korista and collaborators have explored this idea thoroughly in the context of quite generic models with power-law distributions of those various quantities, e.g. \citet{Korista2000, Korista2004, Goad2015}. Such models clearly can produce curved responses, but the degrees of freedom in these generic models is rather large.


On the other hand, the geometry of the BLR in the context of the LOC model could play a significant role in explaining the striking nature of the patterns that we see in MKN~110. \citet{Ferland2020} observed a similar pattern for a single line (He~II $\lambda 4686$) but using a sample of AGN spanning a sufficiently large range of luminosity. They attributed the emission-line deficit observed at the highest luminosities to a BLR with a toroidal geometry of a fixed scale height. If the fact that all three of our studied broad emission lines experience saturation at the highest He~II line luminosities can be explained with such a toroidal geometry, their difference in saturation level could be explained if this BLR torus had a limit in radial extent. Such a limit to the BLR could be naturally provided by the onset of the dusty torus \citep{Netzer1993,Landt2014}. However, this mechanism only works if its inner radius was luminosity-invariant \citep{Pott2010,Landt2019}.    

\subsection{Non-responsive material}

A distributed cloud system can produce a non-linear response, but the observation of differences in response (result R1.3) is most simply explained by the presence of material that does not respond to changes in the ionising flux. One way to achieve this would be to consider multiple thin overlapping clouds, such as in the \cite{Gaskell2009} picture. In this picture, the non-responsive material is nearer the middle. K03b compares mean and rms velocity profiles, and finds that for H$\alpha$ and $H\beta$ the profiles are very similar in shape, suggesting that to first order the non-responsive material is distributed very similarly to the responsive material. The mean H$\alpha$ profile in the K03b data (judged by eye) may be somewhat broader than the rms profile, suggesting that there is more non-responsive material nearer the SMBH, contributing to non-variable emission in the line wings. In MKN~509, spectropolarimetry showed the line wings of H$\alpha$ are more polarised than the line centre, potentially indicating they are dominated by scattered light and therefore have a different behaviour than the line centre \citep[][]{Young1999,Lira2021}. However, for both He~I and He~II, the rms profiles in MRK~110 are clearly broader than the mean profiles, suggesting that the non-responsive He~II material is further out (contributing emission closer to centre of the line).

In a series of papers Devereux and collaborators have argued that some local AGN do not reverberate at all \citep[e.g.,][and references therein]{Devereux2018}, and that therefore BLR emission does not come from clouds, but from a large, optically thin, fully ionised region. Clearly in the MKN~110 case there is reverberating material, but it is possible that we have line flux contributions from both dense clouds and a thin medium in which the clouds are embedded. Rather than being a spherical Str\"{o}mgren-sphere-like arrangement, this idea could fit well into disc-wind models. To go beyond this would need a specific model.

Finally we note that as the response slope changes over historical time (result R1.6), the offset also changes. The non-responsive material therefore also likely has a structural physical response to changes in luminosity, averaged over many years. There is both responsive and non-responsive material, and the amount of both gradually changes. This is an interesting parallel to the conclusion by \citet[][]{Vincentelli2022}, that the BLR contribution to the \textit{continuum}, probably associated with an optically thin component, itself varies over time (on a timescale of months), as a function of the continuum luminosity. Again we see that the BLR evolves on multiple timescales, with structural changes occurring in MKN~110 on timescales from years to decades.

\section{Discussion: Kinematic response in the BLR}
\label{sec:discussion_kinematic_response}

Here we look at the implications of results R2.2, R2.3 and R2.4. As noted in Section~\ref{sec:discussion_timescales}, the complexity of He~II behaviour (result R2.5) is likely because of the short physical timescales relevant to the He~II emitting region, such that its structure is changing almost as fast as the continuum.
This makes it very hard to interpret, and so for the rest of this section we concentrate on the behaviour of H$\beta$. The complexity of the behaviour of He~II will add noise to the correlations among the He~II flux and the parameters for H$\beta$.

Result R2.3, that velocity width decreases as ionising flux increases, is qualitatively consistent with the idea that velocities are in general gravity dominated and so roughly virial, together with the idea that as luminosity increases, clouds further out are lit up. It is hard to go beyond this to a quantitative prediction, as it is likely that the H$\beta$ material covers a wide range of radii, and so a large range of velocities, at any one time.

To explain result R2.2, that velocity offset decreases as flux increases, we need to make an assumption as to what causes that offset. Below, we test several generic possibilities: (1) the offset is caused by a gravitational redshift of the broad line component; (2) the offset is caused by radial \emph{infall} of the BLR material; (3) the offset is caused by a radial \emph{outflow} of BLR material.

\subsection{Gravitational redshift}

If the velocity offset between the narrow and broad lines is caused by gravitational redshift, we can derive the expected relation between redshift and the line velocity width, the latter of which is still set by Keplerian motion. Expressed in terms of the gravitational radius $R_g$, the velocity offset due to gravitational redshift is given by $\Delta v_{gz}=c/\sqrt{R/R_g}$, where R is the radial distance from the SMBH. We can express the virial velocity in the same terms:
$$\sigma_V = \frac{c \sin(i)}{f\sqrt{R/R_g}} = \frac{\sqrt{c \Delta v_{gz}}\sin(i)}{f},$$
where $i$ is the inclination, and $f$ is the virial factor relating observed velocity width to virial velocity (Section~\ref{sec:mkn110_history}). Figure~\ref{fig:mrk110_offset_sigma} shows this prediction for $\sigma_{broad}$ as a function of the offset ($\Delta v_{gz}$ in the expression above). For an inclination of $i=21^\circ$ \citep{Kollatschny2003b} and a virial factor $f=4.3$ \citep[][]{Grier2013},
the model does not reproduce the shape of the correlation.
For simplicity we show the relation for a specific value of $f$, rather than for a range of possible values; however, no value of $f$ results in the right shape to match the data. We are implicitly assuming a single specific radius for the material, but the quadratic relationship should be the same at every radius, so it is not obvious that a distributed system of clouds can produce the right shape. A large mass with a pole on view produces a qualitative explanation for the discrepancy between the lag-based mass estimate and the estimate based on a presumed gravitational redshift; however, it does not fit the shape of the correlation that we find between line width and offset.

\subsection{Infall}

If the motion of the clouds has a net infall component, the near-side clouds appear redshifted, and the far-side clouds are blueshifted. Here we define `near-side clouds' as the clouds that are on the same side of the SMBH as the observer, along the line of sight, and `far-side clouds' as the clouds on the other side. In the case of infall, the near-side clouds are therefore moving away from us. To see a net redshift requires that the emission from the far-side clouds is systematically suppressed from along our line of sight. To investigate the possible configurations, we consider two cases: clouds that are optically thick in the H$\beta$ transition and clouds that are optically thin in H$\beta$. In the optically thick case, the clouds will emit anisotropically, such that the side facing the nucleus is brighter. For almost all geometries this would result in the opposite effect -- far-side clouds will appear brighter, as we are observing the parts of the far-side clouds facing the nucleus.

On the other hand, if the clouds are optically thin, we see them all equally well. Suppressing the far-side clouds could perhaps be achieved by blocking them from our view with the accretion disc \citep[such as found for Arp 151;][]{Pancoast-reverb-P2-2014,Pancoast2018}, provided it extends to large enough radii. The correlation between velocity width and offset then implies that the infall velocity is increasing inwards to the SMBH. This is generally consistent with any gravitational model. To go further requires a model that explains the relationship between rotation and infall.

In summary, for an infall model, clouds need to be optically thin, accelerating inwards, and with the far-side shielded by the disc.

\subsection{Outflow}

For a BLR in outflow, the emission from the far-side clouds is redshifted and the emission from the near-side ones blueshifted. As for the case of infall, if the clouds are optically thin, we see them all equally well. An occulting disc would result in the opposite effect - a net blueshift, as we only see the near-sided clouds. For optically thick, anisotropic clouds, we preferentially see the far-side ones (the parts facing the nucleus), so this picture naturally gives a net redshift as long as there is \textit{not} an occulting disc.  The width-offset correlation then implies that the clouds are decelerating as they move radially outward. 

Outward deceleration is not what would be predicted for a radiation-pressure driven outflow. However, this behaviour could result from a ballistic outflow, which receives an initial impulse nearer the central source and then decelerates under gravity. However, the deceleration would be very fast - the offset velocity changes by a factor of five, while the velocity width changes by about 25\%. This implies the presence of a drag force, as proposed in the `Quasar Rain' model of \citet[][]{Elvis2017}. In this model, a thin disc wind is radiatively accelerated at some radius on the disc, and the BEL clouds are condensations which form within this wind. The clouds have a very different (lower) radiative force multiplier than the surrounding wind, so they stop accelerating, and furthermore suffer a drag force from the wind. In the end, the clouds fall back towards the disc. In this scenario, we could expect to observe both outflow and infall, associated with different clouds. 

\subsection{Rotation with projection effects}

Finally, we note that apparent velocity offsets can be achieved without a radial velocity component, by projection effects. Consider, for example the case where the BELs are condensations in a conical structure which is primarily rotating, even if it also has an outflow component \citep[e.g.,][]{Elvis2000}. If the clouds are optically thick and therefore emitting anisotropically, the line radiation will effectively arise from the inner surface wall of the cone, with the rotational Doppler shift varying with position around the cone wall. However, the $\cos\theta$ projection effect ($\theta$ with respect to our line of sight) will also vary with position around the wall, which can result in a velocity-dependent suppression. The degree of observed shift will depend on both the opening angle of the cone and the tilt angle towards the viewer - it is possible to result in either net redshifts or net blueshifts. The observation that the amount of offset depends on the velocity width suggests that the cone opening angle increases with larger radii, such as proposed in the model of \citet[][]{Elvis2017}. We leave a detailed comparison of this model to future work.

\section{Conclusions}

The strongly variable output and extensive history of optical monitoring campaigns make MKN~110 an excellent test case for our study of the long-term evolution of the BLR response. Our data-set covers 30 year of spectroscopic observations and provides a uniquely detailed view of the evolution of the BLR on this long timescale. We find that using He~II~$\lambda$4686 as a proxy of the FUV continuum provides a promising method to understand and interpret the behaviour of the lower ionisation lines. We observe a levelling off of the responsivity of the ionisation lines at higher He~II~$\lambda$4686 levels. In our two-component model we fit this behaviour as a saturation level of the response. The stratification of the saturation levels in the line response corresponds with the physical stratification of the line species, as evident from RM studies. This suggests we are seeing the effect of key physical processes in the regulation of BLR emission.

In studying the profiles of the He~II~$\lambda$4686 and H$\beta$ lines, we find that the change in velocity offset between the narrow and broad line components does not fully match expectations, if we assume that the offset is caused by a gravitational redshift of the broad-line component. We explore several alternative explanations for the observed evolution, in which the offset is caused by bulk motion of the broad-line emitting gas. We find several generic models that explain the observed behaviour. However, the response of the BLR to a changing continuum is complex, as perhaps best exemplified by the difference in behaviour between He~II~$\lambda$4686 and H$\beta$. This complexity emphasises the need for a more sophisticated, physical BLR model. Such a model lies outwith the scope of this publication. 

The detailed behaviour that we are able to observe, emphasises the analytical power provided by long-term observations, and in particular by repeated spectroscopy, in the study of AGN. The data-set available for MKN~110 is currently rare, both in its duration and its detail. However, we can look forward to the results from new spectroscopic surveys, such as SDSS-V, that will increase the number of objects where we can study the evolution of the BLR using the techniques outlined in this paper.


\section*{Data Availability}

All fluxes, spectra, spectral fitting results, and the code needed to create the figures presented in this work are available at \url{https://github.com/dshoman/MKN110}.

\section*{Acknowledgements}

DH acknowledges support from DLR grant FKZ 50 OR 2003, as well as from a previous Principal's Career Development PhD
Scholarship from the University of Edinburgh.
HL acknowledges a Daphne Jackson Fellowship sponsored by the Science
and Technology Facilities Council (STFC), UK.
This study has made extensive use of the fluxes published in BK99 and K01 and the authors wish to express their gratitude for the inclusion of these data-sets in their entirety in the aforementioned works.
For the FAST spectra, this paper uses data products produced by the OIR Telescope Data
Center, supported by the Smithsonian Astrophysical Observatory. 


\bibliographystyle{mnras}
\bibliography{mkn110} 



\bsp	
\label{lastpage}
\end{document}